\documentclass[11pt]{article}
\usepackage{amsmath, hyperref, wasysym}
\usepackage{latexsym}
\usepackage{cite}
\usepackage{amssymb}
\usepackage{amsthm}
\usepackage{lineno}
\usepackage[margin=0.9in]{geometry}
\usepackage{fancyhdr}
\usepackage{graphicx}
\usepackage{float}
\usepackage{color}
\usepackage{multicol}
\usepackage{authblk}
\usepackage{dsfont}
\usepackage{caption}
\usepackage[T1]{fontenc}
\usepackage{lmodern}
\usepackage{booktabs}
\restylefloat{table}
\begin{document}
\title{Quantifying different modeling frameworks using topological data analysis: a case study with zebrafish patterns}
\author[$\dagger$]{Electa Cleveland}
\author[$\dagger$]{Angela Zhu}
\author[$\dagger$]{Bj\"{o}rn Sandstede}
\author[*]{Alexandria Volkening}
\affil[$\dagger$]{Division of Applied Mathematics, Brown University, Providence, RI, USA (\href{mailto:electa_cleveland@brown.edu}{electa\_cleveland@brown.edu}, \href{mailto:angela_zhu1@brown.edu}{angela\_zhu1@brown.edu}, \href{mailto:bjorn_sandstede@brown.edu}{bjorn\_sandstede@brown.edu})}
\affil[*]{Department of Mathematics, Purdue University, West Lafayette, IN, USA (\href{mailto:avolkening@purdue.edu}{avolkening@purdue.edu})}
\maketitle

\begin{abstract}
Mathematical models come in many forms across biological applications. In the case of complex, spatial dynamics and pattern formation, stochastic models also face two main challenges: pattern data is largely qualitative, and model realizations may vary significantly. Together these issues make it difficult to relate models and empirical data---or even models and models---limiting how different approaches can be combined to offer new insights into biology. These challenges also raise mathematical questions about how models are related, since alternative approaches to the same problem---e.g., cellular Potts models; off-lattice, agent-based models; on-lattice, cellular automaton models; and continuum approaches---
treat uncertainty and implement cell behavior in different ways. To help open the door to future work on questions like these, here we adapt methods from topological data analysis and computational geometry to quantitatively relate two different models of the same biological process in a fair, comparable way. To center our work and illustrate concrete challenges, we focus on the example of zebrafish-skin pattern formation, and we relate patterns that arise from agent-based and cellular automaton models. 
\end{abstract}

\noindent {\small{{\bf{Key Words:}} 
agent-based model, cellular automaton, topological data analysis, alpha-shapes, variability, zebrafish, pattern formation\\

\noindent {\small{{\bf{MSC Codes:}} 
92B05, 92C15, 70G60, 55N31

\section{Introduction}\label{sec:intro}

Whether focused on vegetation patterns \cite{Gowda2014}, schooling and swarming \cite{Katz2011fish,Bernoff2011}, collective cell dynamics \cite{volkeningRev,Buttenschon2020,Blanchard2019,Giniunaite2020,Osborne2017}, or another complex system, researchers use a wide range of mathematical models to study pattern formation. These approaches include stochastic individual-based models, which may operate on or off lattice, and macroscopic models in the form of partial differential equations. This wealth of frameworks is a benefit, since different types of models have complementary strengths and weaknesses, but it also presents challenges. Relating different models of the same biological system is not straightforward, particularly when considering spatial dynamics and detailed, stochastic models, which often rely on qualitative observation to judge model output. Motivated by this challenge, we show how to adapt methods from topological data analysis and computational geometry to quantitatively describe two models of the same biological system in a comparable way. For concreteness, we center our study on zebrafish-skin patterns, and we apply our methods to two stochastic, individual-based models \cite{volkening2018,Owen2020}.

As we show in Figure~\ref{fig:motivation}, zebrafish (\emph{Danio rerio}) are small fish known for their black stripes and gold interstripes \cite{volkeningRev,Parichy2021,IrionRev2019}. These stripe patterns are made up of brightly colored pigment cells and emerge from pigment-cell interactions and the tissue environment \cite{Yamaguchi,Maderspacher2003,ParTur130,Parichy2021}. The diversity of patterns that mutant fish sport when cell interactions are altered makes zebrafish skin an attractive place to study pattern formation mathematically \cite{volkeningRev,Kondo2021rev}. Toward this end, zebrafish patterns have been investigated using many of the modeling approaches that are also used to study biological self-organization in other applications. Microscopic off-lattice \cite{Cops,volkening2015,volkening2018,volkening2020,volkeningThesis} and on-lattice \cite{Bullara,MorDeutsch,Owen2020,Konow2021} models have considered the behavior of individual cells. On the macroscopic side, continuum models \cite{Woolley2017,painter,Bullara,Yamaguchi,Nakamasu,Bloomfield,Gaffney,Konow2021}, including reaction-diffusion equations and integro-differential equations, have been developed to track cell densities. Other approaches \cite{McCalla2018} include reducing complexity by only describing the evolution of the stripe--interstripe interface.

For zebrafish skin and other biological patterns, continuum models have the benefit of being analytically tractable, and they provide a broad perspective on the overarching features (such as density-dependent motion \cite{Cahn,Liu2013mussel} or short-range activation and long-range inhibition \cite{Turing,Mein}) 
that may be at work. On the other hand, agent-based and cellular automaton models\footnote{Researchers use the terms ``agent-based", ``individual-based", and ``cellular automaton" to describe a rich diversity of models. In this manuscript, ``agent-based" means a model that works at the scale of agents---e.g., cells---interacting in continuous space. We distinguish this from what we call ``cellular automaton" models, which also focus on the dynamics of agents, but in which space is discrete; see Figure~\ref{fig:motivation}.}, not analytically tractable using traditional techniques, are often closer to the underlying biology and can make detailed predictions about cell interactions. Uncovering how different types of models are related is a rich mathematical problem, and, because it breaks down silos and allows us to bring the predictions of different studies together, it has important biological applications. To this end, much work has focused on relating macroscopic and microscopic approaches. In the case of zebrafish patterns, for example, Bullara \emph{et al}. \cite{Bullara}, Konow \emph{et al.} \cite{Konow2021}, and Volkening \emph{et al}. \cite{volkening2015,volkeningThesis} developed continuum models for cell density alongside microscopic models. While open questions certainly remain in relating macroscopic and microscopic models, much less work has focused on understanding the relationship between different microscopic models, such as on-lattice and off-lattice approaches. We highlight the studies of Osborne \emph{et al.} \cite{Osborne2017} and Plank \emph{et al.} \cite{Plank2012} on cell behavior as examples.

The increased level of biological detail in individual-based models comes at the cost of more parameters and models rules. The computational complexity of microscopic models---whether they are agent-based, cellular automaton, cellular Potts \cite{Glazier1993,Merks2017}, vertex models, or something else---also leads to more choices and hidden parameters. (See Buttensch\"{o}n \emph{et al}. \cite{Buttenschon2020} for a recent review of computational models of cell behavior.) This raises questions about how choices of computational implementation affect model predictions \cite{Kursawe2017,Osborne2017}. However, the stochastic, complex nature of individual-based models makes it challenging to quantitatively describe---let alone robustly compare and relate---model output. On top of this, qualitative observation of a few example simulations may not capture the breadth of patterns that a stochastic model can generate under the same set of parameters. When models of pattern formation become more and more detailed, they hit more and more of the same challenges faced by the underlying biological data that they represent.

\begin{figure}[t!]
\centering
    \includegraphics[width=0.9\textwidth]{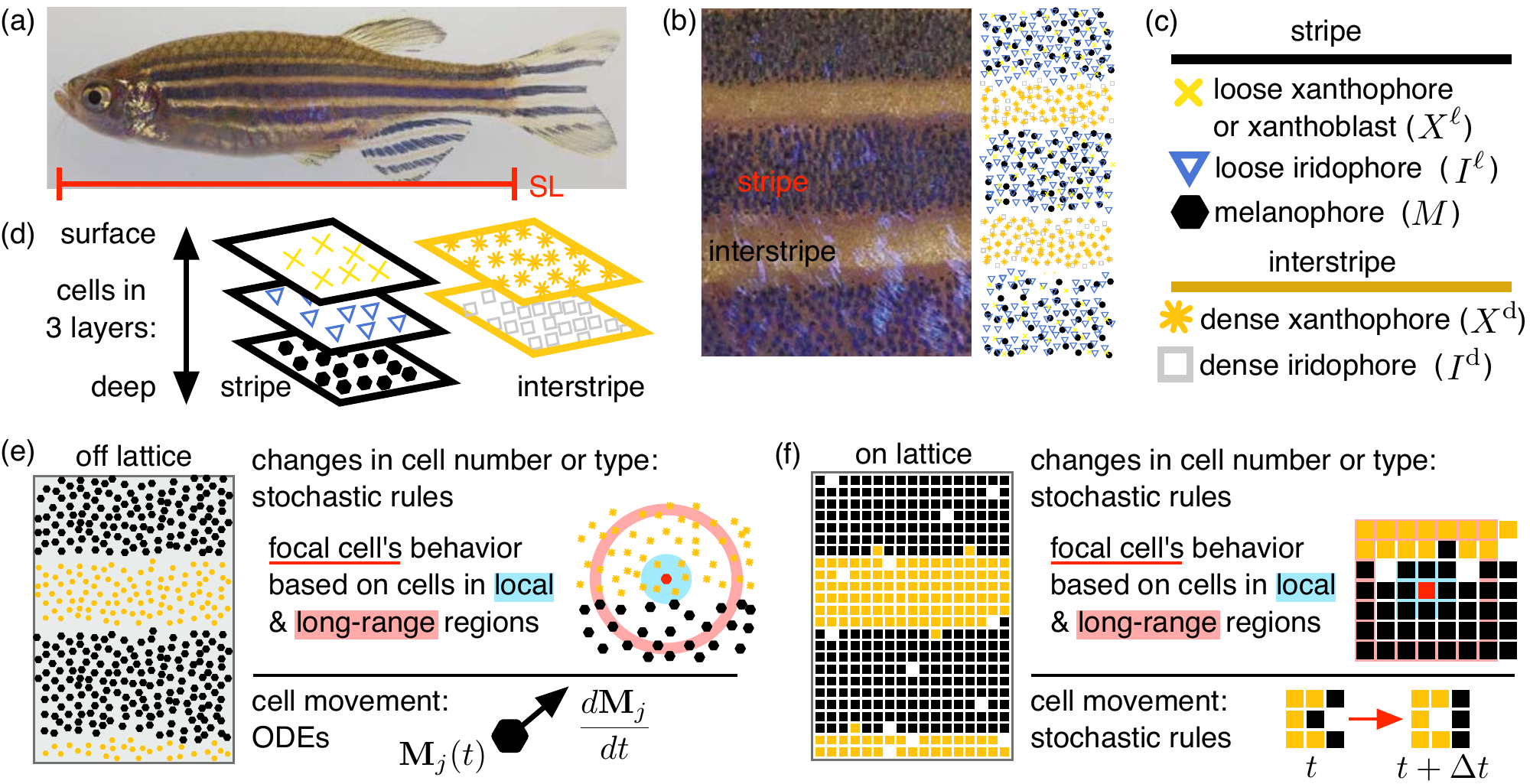}
    \caption{Motivation and background. (a) Zebrafish (\emph{Danio rerio}) are a model organism for studying vertebrate development and skin-pattern formation \cite{IrionRev2019,ParichyRev2019,volkeningRev}. (b) Their stripes and interstripes form due to the interactions of pigment cells \cite{Nakamasu}. (c) Stripes feature $X^\ell$, $I^\ell$, and $M$ cells, and interstripes consist of $X^\text{d}$ and $I^\text{d}$ cells. (d) These cells are organized in three layers, with xanthophores at the surface level \cite{hirata2005pigment,Mcmen2014,Mahalwar,ParTur130}. (e) Prior modeling approaches to pattern formation in zebrafish include stochastic agent-based (off-lattice microscopic) models and cellular automaton (on-lattice microscopic) models \cite{volkeningRev}. Agent-based models \cite{volkening2015,volkening2018,volkening2020,volkeningThesis} treat cells as particles that interact through ODEs for cell movement and stochastic, discrete-time rules for differentiation, division, and death. (f) Cellular automaton models \cite{Owen2020,Bullara,Konow2021} consider space as discrete, and all cell interactions are stochastic rules. Both of these approaches are not analytically tractable using traditional techniques, and they often involve qualitative observations to judge model output. This makes it difficult to relate different models, and this challenge motivates our study. Images (a, b) adapted from Fadeev \emph{et al.} \cite{fadeev2015} and licensed under CC-BY $4.0$ (\url{https://creativecommons.org/licenses/by/4.0/}); we added the SL bar, text, and schematic in (a, b); images (c, d) adapted from Volkening \emph{et al}. \cite{volkening2018} under CC-BY 4.0; images (e, f) adapted from Volkening \cite{volkeningRev} with permission from Elsevier, Copyright (2020) Elsevier Ltd.
    \label{fig:motivation}}
\end{figure}

Whether studying zebrafish patterns or another biological system, the first step to relating modeling frameworks at large scale is quantitatively describing self-organization. Methods for quantifying \emph{in vivo} and \emph{in silico} patterns include pair-correlation functions \cite{Dini2018,Treloar2014spatial,Gavagnin2018yates}, order parameters \cite{Couzin2002,Chuang2007,Huepe2008}, and pattern-simplicity scores \cite{Miyazawa2010Simplicity,Miyazawa2020}. Recently tools from topological data analysis---particularly persistent homology---\cite{Edelsbrunner2008,Carlsson2009, Carlsson2020, Otter2017mason,Ghrist2014,Carlsson2005,chazal,Munch2017} have also emerged as a flexible means of quantifying complex systems and pattern formation \cite{Topaz2015,Ulmer2019topaz,Munch2020Shape}. Persistent homology has been applied to many biological systems, including flocking \cite{Topaz2015,Bhaskar2019}, vascular networks \cite{Nardini2021arxiv,Bendich2016,Stolz2022}, migrating cells \cite{Bonilla2020,Bhaskar2021}, intracellular dynamics \cite{Ciocanel2021}, and fish skin \cite{McGuirl2020}. 

Persistent homology characterizes connected components, holes, and higher-dimensional features in data across scales \cite{Otter2017mason}, and interpreting topological summaries can involve choices and hyper-parameters that are application specific. Recently McGuirl \emph{et al.} \cite{McGuirl2020} developed a pipeline for interpreting information from persistent homology as descriptions of spots and stripes in an agent-based model of zebrafish patterns \cite{volkening2018}. The quantification results \cite{McGuirl2020} involve several hyper-parameters, which are motivated by length scales in the model \cite{volkening2018}. A related issue is that persistent homology is sensitive to outliers \cite{Fasy2014,Bendich2011}, and this is challenging in biological settings, where noise is inherent. This raises questions about how McGuirl \emph{et al.}'s approach can be adapted to quantify zebrafish patterns in noisier \emph{in vivo} settings or alternative \emph{in silico} settings, and about how sensitive the methods \cite{McGuirl2020} are to the choice of hyper-parameters when interpreting topological summaries.

Motivated by the challenges associated with meaningfully relating different stochastic, microscopic models, here we apply methods from topological data analysis and computational geometry to quantify stripe patterns from two models of zebrafish at large scale. The two focal models of our case study---one off-lattice \cite{volkening2018} and one on-lattice \cite{Owen2020} individual-based model---are a useful place to center our work because they have largely the same cell interactions at their core from a biological perspective, but their computational implementation, cell density, and noise structure differ significantly. Drawing on these differences and similarities, we build on the methods \cite{McGuirl2020} to choose hyper-parameters in a way that ensures our comparisons are robust and fair. We also apply $\alpha$-shapes, a tool from computational geometry \cite{Esurvey,Edelsbrunner1983}, as a flexible means of identifying interstripe boundaries and computing curviness. Our work illustrates the challenges that arise when relating biologically similar---but mathematically and computationally different---stochastic models, and it shows how choices in the quantification process affect interpretations of results. This helps open the door to future large-scale studies of stochastic, spatial models, centered on questions including how sensitive model predictions are to implementation choices, and on how alternative mathematical perspectives can be combined to address biological questions.

\section{Background and methods}\label{sec:background}

We first provide a brief overview of zebrafish-pattern biology (\S \ref{sec:biology}) and the two models \cite{volkening2018,Owen2020} that serve as the basis of our case study (\S \ref{sec:models}). Because our quantification pipeline relies on tools from topological data analysis and computational geometry, we then introduce persistent homology \cite{McGuirl2020,Otter2017mason} and $\alpha$-shapes \cite{Esurvey,Edelsbrunner1983} in \S \ref{sec:tda}.

\begin{figure}[t!]
\centering
    \includegraphics[width=0.9\textwidth]{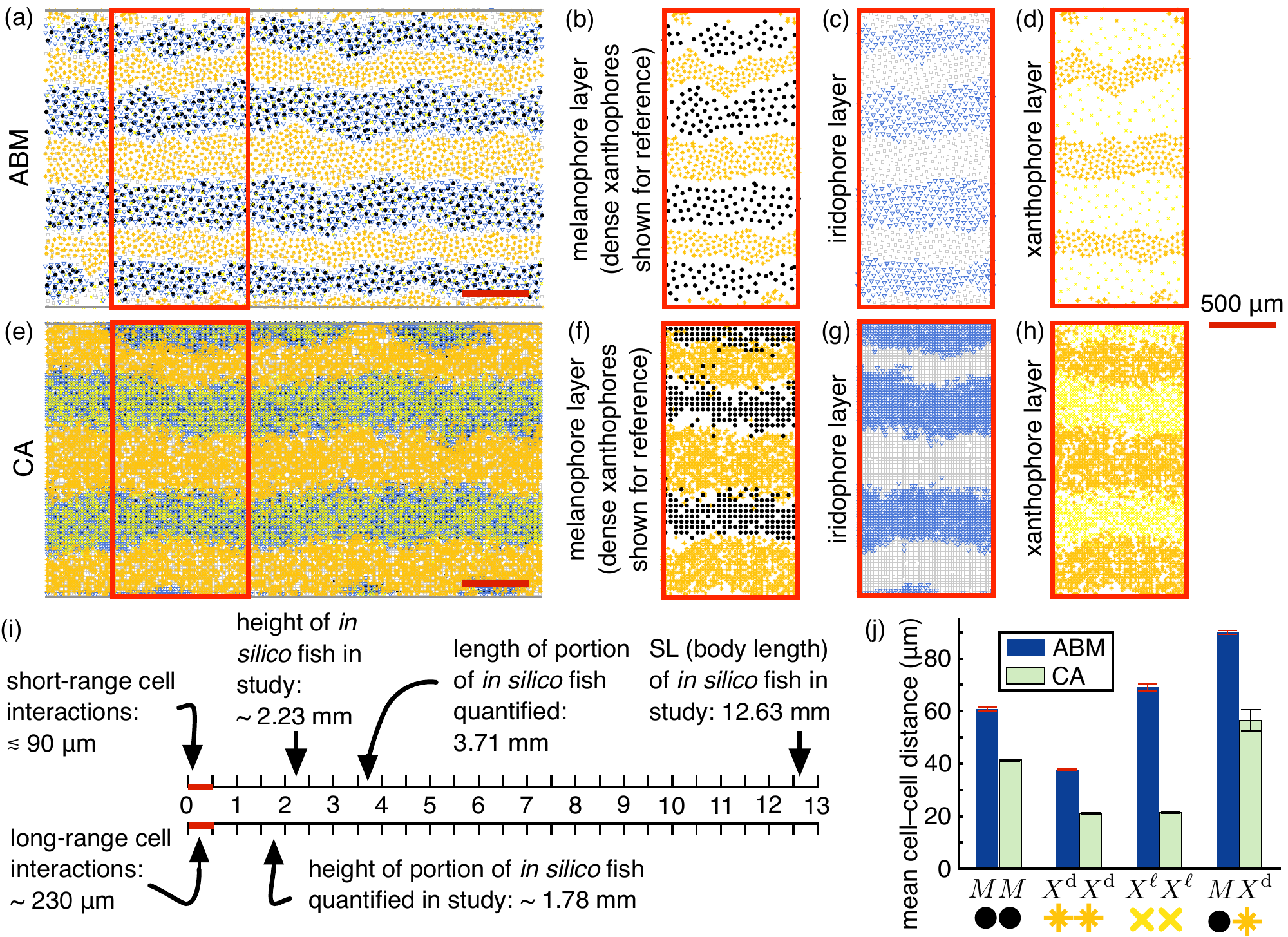}
    \caption{Patterns and length scales for our focal models \cite{volkening2018,Owen2020}.  (a) The ABM \cite{volkening2018} produces patterns with interstripe and stripe cells in distinct regions. (b) Cells occupy three layers in zebrafish skin (Figure~\ref{fig:motivation}(d)) \cite{hirata2005pigment}, and we show these layers separately for easier viewing. (We include $X^\text{d}$ cells with $M$ cells for reference.) 
    In the ABM \cite{volkening2018}, the presence of three layers of cells is taken into account indirectly by specifying volume-exclusion forces only between cells that occupy the same layer of skin. 
    Orange $X^\text{d}$ cells are separate from $M$ cells. (c) Blue $I^\ell$ and silver $I^\text{d}$ cells occupy different regions in the ABM, as do (d) yellow $X^\ell$ and orange $X^\text{d}$ cells. (e) CA \cite{Owen2020} patterns feature xanthophores and iridophores at higher density than ABM patterns, and (f) there is some overlap of $M$ and $X^\text{d}$ regions, (g) $I^\ell$ and $I^\text{d}$ regions, and (h) $X^\ell$ and $X^\text{d}$ regions. For comparison, we use the same symbols to plot cells in the ABM and CA patterns in (a)--(h), highlighting differences in cell density. (i) Length scales in our study range from less than $90$ $\mu$m to over $10$ mm \cite{volkening2018,Owen2020,ParTur256,Parichy,TakahashiMelDisperse,2016heterotypic}. (j) We show the mean distance between nearest-neighboring cells, averaging across $1000$ simulations \cite{volkening2018,Owen2020}. (The distance between $M$ and $X^\text{d}$ cells is for cells at stripe--interstripe boundaries: to approximate this, we take the mean distance between nearest $M$ and $X^\text{d}$ cells with distances greater than or equal to $110$ $\mu$m or less than $0$ $\mu$m excluded. This choice may overestimate the distance between $M$ and $X^\text{d}$ cells for the CA \cite{Owen2020} slightly, since these cells can occupy the same position in different grids: if we consider distances of $0$~$\mu$m, the mean distance between $M$ and $X^\text{d}$ cells decreases by about $2$~$\mu$m.) Bars indicate the standard deviation in the mean distances.}\label{fig:models}
\end{figure}

\subsection{Biological background} \label{sec:biology}

Wild-type zebrafish are characterized by dark stripes and gold interstripes that form in their skin \cite{IrionRev2019,ParichyRev2019,volkeningRev}. These patterns emerge from the interactions of tens of thousands of pigment cells arranged in three layers \cite{hirata2005pigment} (Figure~\ref{fig:motivation}(d)). There are three main types of cells involved in patterning: black melanophores ($M$), dense orange ($X^\text{d}$) or loose yellow ($X^\ell$) xanthophores, and dense silver ($I^\text{d}$) or loose blue ($I^\ell$) iridophores \cite{volkeningRev}. The melanophores---present in the deepest layer---are restricted to dark stripes on the main fish body, whereas the other two cell types appear across the pattern in different colors and forms \cite{hirata2005pigment,Mcmen2014,ParTur130,Mahalwar,Frohnhofer,Hirata}. Housed in the surface layer, $X^\text{d}$ cells are tightly packed in interstripes, and $X^\ell$ cells (also called xanthoblasts) appear in a looser form at lower density in stripes \cite{Mahalwar}. Similarly, iridescent iridophores are present as $I^\text{d}$ in interstripes and $I^\ell$ in stripes, sandwiched in the middle layer between melanophores and xanthophores in the skin \cite{hirata2005pigment,Singh}.

As wild-type zebrafish develop over a few months, all of these pigment cells interact through movement, division, death, and differentiation to form patterns. Cell–cell interactions may be mediated by diffusing signals \cite{Patterson2014}, cellular extensions \cite{Inaba,delta,eom2017macrophage}, or other mechanisms; in all cases, cells occupy a messy, biological environment, and there is inherent noise in their behavior. New pigment cells can appear through division of existing cells or by differentiation of precursors \cite{Gur2020,Mahalwar,Mcmen2014,Singh,Budi,Dooley}. Some cell behaviors are relatively well understood, and in other cases the interactions underlying cell dynamics are unknown \cite{2016heterotypic,kondoTuringQuestion}. Iridophores, in particular, are a place in zebrafish-pattern biology where the field has been actively evolving \cite{Singh,PatPLos,Patterson2014,Jan,2016heterotypic,Gur2020}. Up until roughly 2014, the focus was on melanophores and xanthophores, and iridophores were not heavily considered (e.g., \cite{TakahashiMelDisperse,Maderspacher2003,Yamaguchi,ParTur130}). This changed with new observations \cite{Singh} suggesting iridophores may swap their shape and color in specific ways (e.g., a loose blue iridophore transitioning to a silver iridophore). More recently empirical understanding of iridophores has continued to grow and shift, with Gur \emph{et al.} \cite{Gur2020} suggesting in late 2020 that iridophores differentiate in dense or loose form from precursors, rather than changing their shape once established.

Because our work applies persistent homology to understand cell organization, the length scales involved in zebrafish patterns are particularly important in this study. Roughly, the distance between nearest-neighboring cells tends to be around $50$ micrometers ($\mu$m) \cite{TakahashiMelDisperse,ParTur256,2016heterotypic}, and, at the pattern scale, stripes and interstripes in adult fish are about $500$~$\mu$m wide \cite{volkening2018,delta,fadeev2016}; see Figure~\ref{fig:models}(i). As we discuss in \S \ref{sec:models}, the simulated patterns that we quantify correspond to pre-adult fish that are $12.63$ millimeters (mm) long in ``standard length" (SL), which refers to the distance from fish snout to where the body and tailfin meet (Figure~\ref{fig:motivation}(a)) \cite{Parichy}. Because zebrafish grow at different rates, it is customary to describe age in terms of standardized standard length (SSL, a measurement of SL associated with a reference zebrafish) or developmental stage, rather than time in days post fertilization \cite{Parichy}. When they reach $12.63$~mm SSL, zebrafish are in the juvenile developmental stage; at this point, patterns generally include three light interstripes and two dark stripes. Measuring from the front edge of the bottom fin in Figure~\ref{fig:motivation}(a), this corresponds to a fish body that is about $2.3$ mm high \cite{Parichy}.

\subsection{Focal mathematical models} \label{sec:models}

Our case study centers on two closely related models that are stochastic and biologically detailed: the off-lattice model \cite{volkening2018} developed by Volkening and Sandstede in 2018, and the on-lattice model \cite{Owen2020} built by Owen, Kelsh, and Yates in 2020. For the remainder of this manuscript, we refer to the former as the ABM \cite{volkening2018} and the latter as the CA \cite{Owen2020}; see Figure~\ref{fig:models}. In terms of the wild-type cell interactions that they describe and predict, the models \cite{volkening2018,Owen2020} are largely the same from a biological perspective. They differ primarily in choices of computational implementation, which, in turn, affect how stochasticity shows up in cell behavior and pattern features. (We focus on the models \cite{volkening2018,Owen2020} because of their biological similarities and mathematical differences; in the future, it would be interesting to consider other models that differ in the types of cells that they consider, such as \cite{Konow2021,volkening2015,Bullara,Shinbrot}.) Both models \cite{volkening2018,Owen2020} were built pre-publication of \cite{Gur2020}, in a biological environment in which the prevailing view was that iridophores may undergo shape and color changes. Because our focus is on developing methods for comparing similar models implemented in different ways, the new findings \cite{Gur2020} do not affect our study. We overview the ABM and CA here; see Appendix \ref{sec:app} for a deeper dive into a few example model rules, and \cite{volkening2018,Owen2020} for full details.

The ABM \cite{volkening2018} and CA \cite{Owen2020} both include five cell types: black $M$, blue $I^\ell$, and yellow $X^\ell$ cells in stripes, and silver $I^\text{d}$ and orange $X^\text{d}$ cells in interstripes. They also account for five dynamics: fish growth, along with cell movement, division or differentiation, death, and shape changes. The ABM \cite{volkening2018} treats cells as particles interacting in continuous space, and tracks the positions (i.e., $(x,y)$-coordinates) of cells (point masses) interacting in growing domains. Volkening \emph{et al.} implement cell movement through coupled differential equations: each pigment cell has an ODE describing the forces that it feels from other cells. Cell differentiation, division, death, and changes in shape---the dynamics that influence cell number---all take the form of stochastic, discrete-time rules. The ABM \cite{volkening2018} accounts for three layers of cells in 2D domains indirectly by specifying volume-exclusion forces only between cells that occupy the same layer in the skin. On the other hand, the CA \cite{Owen2020} is discrete in space, allowing cells to occupy positions in grids which grow in time through the addition of new grid squares. All cell behaviors---movement, differentiation, division, death, and shape changes---in Owen \emph{et al.}'s model are based on stochastic, continuous-time rules, and they \cite{Owen2020} include three separate 2D grids for melanophores, xanthophores, and iridophores.

Both approaches \cite{volkening2018, Owen2020} simulate pattern formation starting from the pelvic-bud developmental stage, when zebrafish are about $7.6$ mm long in standard length and roughly $1.0$ mm high. Volkening \emph{et al.} \cite{volkening2018} and Owen \emph{et al.} \cite{Owen2020} start with a rectangular domain that represents the full fish height ($1$ mm) and a portion of the fish body length. Because the models account for the remainder of the fish body differently, their domains grow at different rates. As we discuss in \S \ref{sec:step1}, we focus on quantifying patterns that correspond to juvenile zebrafish at $12.63$ mm in standard length, and this milestone is associated with ABM and CA patterns on domains of different sizes. Growth affects cell positions in both models: in the ABM, cell positions are multiplied by a scaling factor to account for uniform growth in a spatially continuous manner. In the CA, grid squares are inserted at randomly selected locations, modeling uniform growth in a spatially discrete way.

In addition to treating space as continuous or discrete, the models \cite{volkening2018,Owen2020} differ in how they treat time. The ABM has a time step of one day, and three main dynamics occur on each simulated day of fish development. First, the domain size is increased and cell positions are stretched to account for fish growth in a deterministic manner. Second, all of the cells undergo deterministic migration simultaneously; and third, the numbers and types of cells present are simultaneously updated to account for cell death, differentiation, division, and shape changes. For example, this means that every existing $M$ cell is evaluated simultaneously for possible death (according to stochastic rules) each day. In comparison, the CA functions in continuous time, with an exponential clock that goes off when an ``event" should occur. This event could be vertical domain growth, horizontal domain growth, evaluating a randomly selected $M$ cell for potential death, etc. Owen \emph{et al.} \cite{Owen2020} selected most rates for each of these events so that all of their cells are evaluated for that event once per day. This suggests that both models may be capturing the same number of behaviors per day on average. However, the ABM implements cell interactions simultaneously and the CA updates cell populations iteratively.

\subsection{Geometric and topological data analysis} \label{sec:tda}

The microscopic models \cite{volkening2018,Owen2020} that serve as the basis of our case study produce point-cloud data in the form of pigment-cell locations. As we show in Figure~\ref{fig:models}, stripes are visible in patterns generated by these models \cite{volkening2018,Owen2020}, but the discrete nature of the data makes them challenging to quantitatively describe. To help address this, our methodology for quantifying stripe patterns in \S \ref{sec:methods} draws on techniques from topological data analysis and computational geometry. Specifically, we utilize persistent homology and $\alpha$-shapes, which naturally lend themselves to point-cloud data. We provide an informal introduction to persistent homology and $\alpha$-shapes here; see \cite{Topaz2015,Otter2017mason,Edelsbrunner2008,Carlsson2009,chazal,Ghrist2014} and \cite{Esurvey,Edelsbrunner1983}, respectively, for more technical details.

\subsubsection{Persistent homology}\label{sec:PH}

Persistent homology provides information about the number of connected components, holes, trapped volumes, and higher-dimensional features in data across scales \cite{Otter2017mason}. Topaz, Ziegelmeier, and collaborators \cite{Topaz2015, Ulmer2019topaz, Bhaskar2019} have used persistent homology to understand the shape of complex-systems data, and topological techniques are being increasingly applied to \emph{in vivo} and \emph{in silico} biological systems \cite{Munch2020Shape}; examples include \cite{Bhaskar2019,Bonilla2020,McGuirl2020,Ciocanel2021,Nardini2021arxiv}. Simplicial complexes are at the core of persistent homology, and we focus on one type of simplicial complex, namely the Vietoris--Rips complex. To build a Vietoris--Rips complex, we start with point-cloud data, such as the positions of $N$ cells in a domain. We place a ball with diameter $\varepsilon$ around each point and connect two points with an edge whenever their respective $\varepsilon/2$-balls intersect, indicating that the two points are at most a distance $\varepsilon$ apart \cite{Topaz2015,Bhaskar2019}. If three points are all at most $\varepsilon$ apart, we note this by filling in the triangle bound by the edges between pairs of points. Data points are $0$-simplices, edges connecting two points are $1$-simplices, and filled-in triangles are $2$-simplices \cite{Topaz2015,Bhaskar2019}. We can continue this process to generate $k$-simplicies for $k \ge 0$, but we focus on $k=0$ and $k =1$ here.

\begin{figure}[t!]
\centering
    \includegraphics[width=0.9\textwidth]{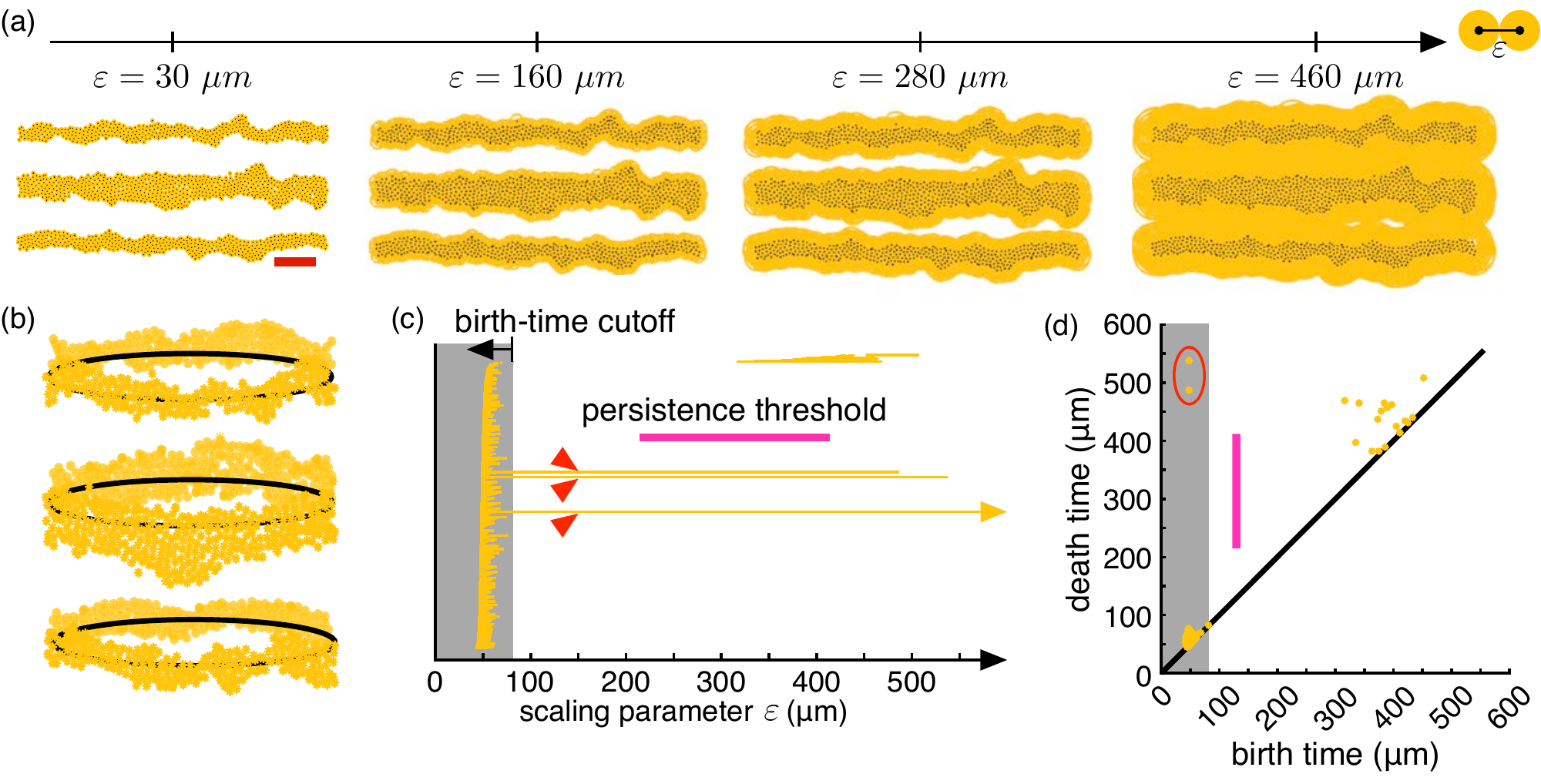}
\caption{Persistent homology of zebrafish patterns \cite{McGuirl2020}. (a) As an example, we show gold balls of varying diameter $\varepsilon$ centered at the positions of $X^\text{d}$ cells (black points here) from the ABM \cite{volkening2018}. (b) Boundary conditions are periodic in $x$ for the models \cite{volkening2018,Owen2020}, so unbroken stripes and interstripes form loops. (c) We show the barcode describing the dimension-$1$ topological features associated with the pattern in (a). McGuirl \emph{et al.} \cite{McGuirl2020} interpreted bars with birth times less than a birth-time cutoff ($80$ $\mu$m) and persistence greater than a threshold ($200$ $\mu$m) as corresponding to interstripes. There are three long bars (marked by red triangles) that meet these conditions, corresponding to three clear interstripes in (a). (d) Persistence diagrams are an alternative means of summarizing information from persistent homology; here each point represents the birth and death times of a loop. Because the final loop in (a)---a function of our periodic domain---persists longer than the maximum $\varepsilon$ value that we considered, only two persistent loops (circled in red) are visible.}
    \label{fig:TDA}
\end{figure}

By varying $\varepsilon$, we can study how the shape of our point-cloud data evolves across scales. Specifically, we are interested in the presence of $k$-dimensional holes, meaning empty regions bound by $k$-simplices \cite{Bhaskar2019,Topaz2015}. Connected components are $0$-dimensional holes, and $2$D trapped areas are $1$-dimensional holes (i.e., loops) bound by $1$-simplices. As we allow $\varepsilon$ to grow, the shape of our data in dimension $0$ evolves from $N$ isolated points when $\varepsilon = 0$ to one large connected component once $\varepsilon$ is large enough. McGuirl \emph{et al}. \cite{McGuirl2020} interpreted dimension-$1$ topological features, in particular, as holding information about stripe patterns generated by the ABM \cite{volkening2018}. Because the ABM \cite{volkening2018} has periodic boundary conditions in $x$, pattern formation is occurring on a cylinder, as we show in Figure~\ref{fig:TDA}(b). If we apply persistent homology to the positions of $X^\text{d}$ cells on this cylindrical domain, interstripes are visible as loops in our data for a range of $\varepsilon$ values. On the other hand, applying persistent homology to the positions of $M$ or $X^\ell$ cells translates into information about stripes \cite{McGuirl2020}.

To make this relationship between (inter)stripes and dimension-$1$ topological features more precise \cite{McGuirl2020}, it is useful to discuss birth and death times \cite{Topaz2015}. The ``birth time" of a topological feature is the value $\varepsilon = \varepsilon_\text{birth}$ at which the feature first appears, and the ``death time" is the value $\varepsilon = \varepsilon_\text{death}$ at which the feature disappears \cite{McGuirl2020}. A feature's ``persistence" is $\varepsilon_\text{death}-\varepsilon_\text{birth}$, and this information is often visualized using barcodes and persistence diagrams. We show the barcode and persistence diagram associated with dimension-$1$ topological features for an example of three interstripes in Figure~\ref{fig:TDA}. 
In a barcode, we draw a horizontal bar starting at $\varepsilon_\text{birth}$ and ending at $\varepsilon_\text{death}$ for each topological feature \cite{Topaz2015}; these bars are stacked vertically. A persistence diagram is a scatter plot with birth time as the $x$-axis and death time as the $y$-axis; each point $(\varepsilon_\text{birth},\varepsilon_\text{death})$ refers to the birth and death times of a feature. As we show in Figure~\ref{fig:TDA}(c), there are three long bars, corresponding to three highly persistent loops. Specifically, McGuirl \emph{et al.} \cite{McGuirl2020} defined the number of interstripes in ABM \cite{volkening2018} patterns by counting the number of $X^\text{d}$ loops with persistence greater than a persistence threshold, and birth time less than a birth-time cutoff. The persistence threshold and birth-time cutoff are hyper-parameters. In \S \ref{sec:step3}, we build on the work \cite{McGuirl2020} to make this approach more flexible for on-lattice stripe patterns and reliably choose hyper-parameter values.

\subsubsection{Alpha-shapes} \label{sec:alpha}

\begin{figure}[t!]
\centering
\includegraphics[width=0.74\textwidth]{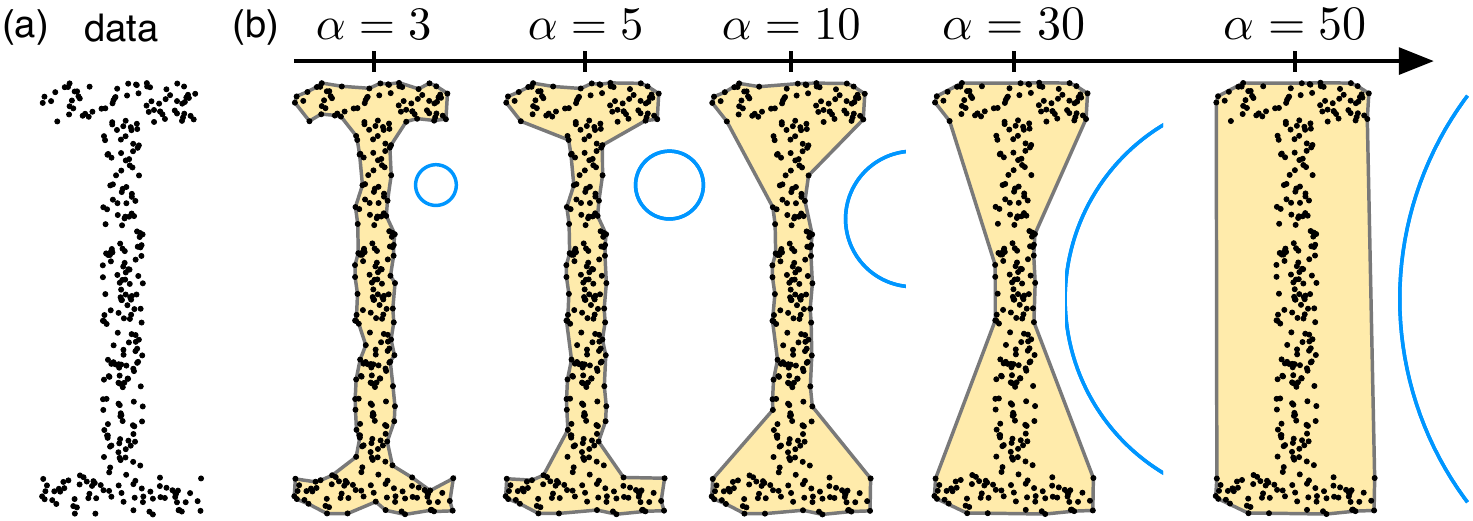}
\caption{Introduction to $\alpha$-shapes \cite{Esurvey,Edelsbrunner1983}. (a) As an example, we consider point-cloud data in the shape of a dumbell turned on its side. (b) Alpha-shapes \cite{Edelsbrunner1983,Esurvey}, a generalization of the convex hull, are a flexible means of characterizing the boundary of point-cloud data. As $\alpha \ge 0$ gets larger, we estimate the boundary of data less and less tightly, approaching the convex hull as $\alpha \rightarrow \infty$. We show the circle (or portion of the circle) with radius $\alpha$ used to construct each $\alpha$-shape in blue. \label{fig:alphaShape}}
\end{figure}

Complementing information about the number of connected components and holes in data that persistent homology can provide, $\alpha$-shapes, a concept from computational geometry, are a flexible means of characterizing the overall shape of data \cite{Esurvey}. A generalization of the convex hull rigorously defined by Edelsbrunner \emph{et al}. \cite{Edelsbrunner1983}, $\alpha$-shapes have been applied to reconstruct surfaces \cite{Guo,Xu}, as well as estimate perimeter \cite{Castro2017}, surface curvature \cite{Li2012}, shape complexity \cite{Gardiner2018}, and volume \cite{Liang}. For example, $\alpha$-shapes have been used in structural biology \cite{Liang} to better understand the shape of protein structures \cite{Zhou2012}. Like persistent homology, $\alpha$-shapes lend themselves to point-cloud data (i.e., cell coordinates in our case study) and provide information about shape. They also naturally involve the choice of a hyper-parameter $\alpha$, and this inherent flexibility makes them suitable for our study.

Alpha-shapes are often described intuitively using circular erasers or spoons with radius $\alpha$ \cite{Esurvey,Edelsbrunner1983}. In this analogy \cite{Esurvey,Edelsbrunner1983}, constructing an $\alpha$-shape is thought of as carving away from the convex hull of a point cloud whenever it is possible to do so without removing any data point. At the final stage, all of the eraser marks with curvature $1/\alpha$ are approximated as straight edges. In particular, given some set of points, an associated $\alpha$-shape is made up of points and edges representing the boundary of the point cloud \cite{Edelsbrunner1983}. Depending on the value of the hyper-parameter $\alpha$, we can select how tightly the $\alpha$-shape fits our data. For finite $\alpha > 0$, the associated $\alpha$-shape is built by drawing an edge between two data points whenever we can place a ball of radius $\alpha$ in such a way that (1) the open ball contains no points in our data set, and (2) the two data points in question lie on the boundary of the closed ball \cite{Esurvey}.

In Figure~\ref{fig:alphaShape}(a), we show an example point cloud, consisting of discrete points in the shape of an overturned dumbell. When $\alpha$ is large (e.g., $\alpha \approx 50$ in Figure~\ref{fig:alphaShape}), the boundary of the $\alpha$-shape is nearly the convex hull of our data, and we miss meaningful structure in the dumbell shape. Indeed, when $\alpha = \infty$, the $\alpha$-shape is simply the convex hull of the point cloud. As we decrease $\alpha$ in Figure~\ref{fig:alphaShape}(b), our $\alpha$-shapes look more like hourglasses and eventually pick up the shape of the dumbell between $\alpha = 10$ and $\alpha = 5$. When $\alpha = 0$, the $\alpha$-shape is just the data itself \cite{Edelsbrunner1994}. (Notably, $\alpha$-shapes are also defined for $\alpha < 0$; see \cite{Edelsbrunner1983,Edelsbrunner1994} for more information.) While we do not discuss it further here,  there is a connection between $\alpha$-shapes, Voronoi diagrams, and Delaunay triangulations \cite{Liang,Edelsbrunner1983,Esurvey}. Additionally, $\alpha$-shapes are related to $\alpha$-complexes, which---like the Vietoris--Rips complex that we introduced in \S \ref{sec:PH}---can serve as the basis for computing persistent homology \cite{Otter2017mason,Esurvey,Edelsbrunner1994}. In \S \ref{sec:step5}, we apply $\alpha$-shapes to extract stripe--interstripe boundaries and quantify pattern curviness.

\section{Results: Our methodology for quantifying stripe patterns}
\label{sec:methods}

\begin{figure}[t!]
\centering
    \includegraphics[width=\textwidth]{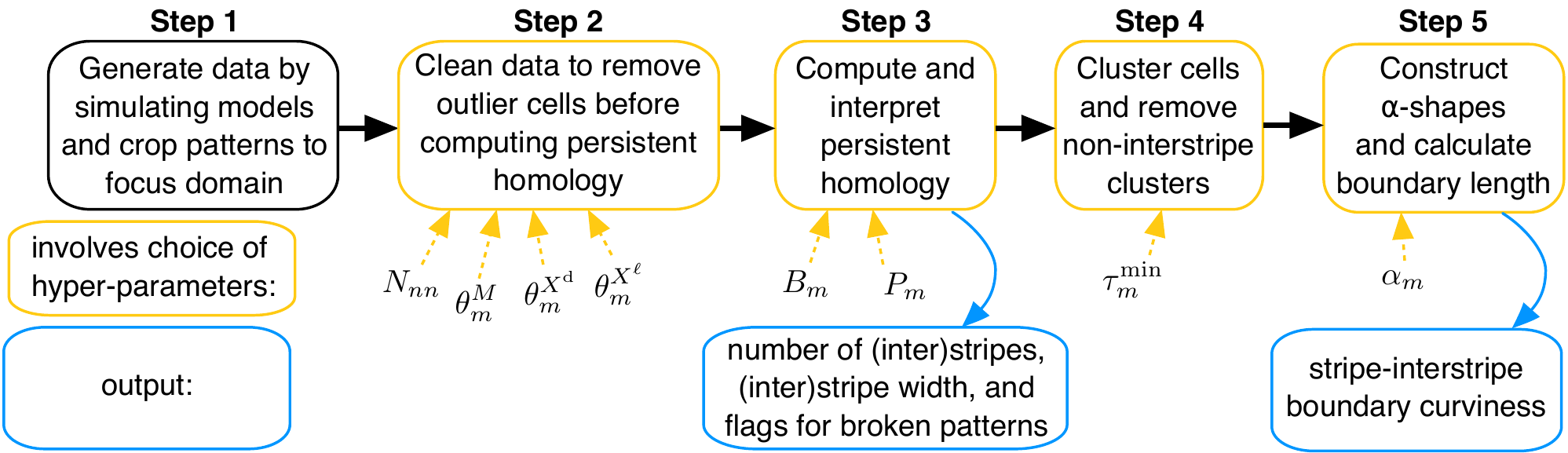}
  \caption{Summary of our quantification process and hyper-parameters. To generate our data, we simulate both models \cite{volkening2018,Owen2020} and crop the final patterns to a focus domain length. Because persistent homology can be sensitive to outliers \cite{Fasy2014,Bendich2011} and CA patterns \cite{Owen2020} often features stray cells, we perform an initial cleaning step before computing persistent homology using the distances between cells on a domain that is periodic in $x$ (see Figure~\ref{fig:TDA}(c)). We use the birth times and persistence of dimension-$1$ topological features---corresponding to loops---to identify the number of stripes and interstripes in each pattern, flag broken patterns, and estimate the maximum width of (inter)stripes \cite{McGuirl2020}. Before constructing $\alpha$-shapes \cite{Esurvey,Edelsbrunner1983} and extracting their boundaries to quantify interstripe curviness, we cluster $X^\text{d}$ cells into the number of interstripes that we identified in Step $3$. If a cluster has fewer than $\tau^\text{min}_m$ cells, it is likely a stray cell or spot, and we remove it and re-cluster, further cleaning the data. Importantly, hyper-parameters enter our methods in Step $2$ (e.g., defining ``outlier"), Step $3$ (i.e., our birth-time cutoff multiplier and persistence threshold), Step $4$ (i.e., $\tau^\text{min}_m$), and Step $5$ (i.e., our alpha radius). As we discuss in \S \ref{sec:step2}, we use $N_{nn}=10$ cells, $\theta^{M}_\text{ABM} = \theta^{M}_\text{CA} = 220$ $\mu$m, $\theta^{X^\text{d}}_\text{ABM} = \theta^{X^\text{d}}_\text{CA} = 130$~$\mu$m, $\theta^{X^\ell}_\text{ABM} = 215$~$\mu$m, and $\theta^{X^\ell}_\text{CA} = 170$~$\mu$m for all of our results. Based on our robustness study in \S \ref{sec:results1}, we set $B_\text{ABM}=B_\text{CA} = 2.5$ mean cell--cell distances as our baseline birth-time cutoff; $P_\text{ABM}=P_\text{CA} = 200$ $\mu$m as our persistence threshold; $\tau^\text{min}_\text{ABM}=\tau^\text{min}_\text{CA} = 60$ cells as our minimum cluster size; and $\alpha_\text{ABM}=\alpha_\text{CA}=100$ $\mu$m as our alpha radius. We use these baseline hyper-parameter values for our results in \S \ref{sec:results2}.}
    \label{fig:pipeline}
\end{figure}

Here we develop a pipeline for quantifying stripe patterns from the off-lattice model \cite{volkening2018} and the on-lattice model \cite{Owen2020} introduced in \S \ref{sec:models}. Our methodology involves cleaning the pattern data to remove outlier cells, computing persistent homology using cell coordinates, building on the methods \cite{McGuirl2020} to interpret persistent homology more flexibly, and constructing $\alpha$-shapes to characterize curviness. Summarized in Figure~\ref{fig:pipeline}, our five-step process allows us to quantify stripe patterns emerging from stochastic on- and off-lattice models in a robust, comparable way. We show how to characterize the number of patterns with broken stripes or interstripes, stripe and interstripe width, and interstripe curviness, and we pay particular attention to pointing out hyper-parameters throughout our process.

\subsection{Step 1: Generate and format data}\label{sec:step1}

To generate the data for our case study on zebrafish patterns, we simulate the ABM \cite{volkening2018} and CA \cite{Owen2020} under wild-type conditions. This step involves mild post-processing to frame the output of both models in the same way. As we discussed in \S \ref{sec:biology}, zebrafish development is generally measured in fish length, rather than in time, so we simulate these models until a standard length of $12.63$ mm. Because the ABM \cite{volkening2018} includes deterministic domain growth, this corresponds to simulating the model for $45$ days, until the fish reaches an age of $66$ days post fertilization. In contrast, the CA \cite{Owen2020} features stochastic domain growth, so there is not a direct correspondence between simulation time, domain size, and fish standard length. Instead, we simulate the CA \cite{Owen2020} until the domain reaches a size associated with a standard length of $12.63$ mm\footnote{Because the CA model \cite{Owen2020} works on a discrete domain, the final domain size is within one grid square (maximally $40$ $\mu$m) of the target length.}. We simulate both models $1000$ times and save the results at the final time.

The positions of cells in the ABM \cite{volkening2018} are in the form of $(x,y)$-coordinates in $\mu$m, while the positions of cells in the CA \cite{Owen2020} are described using lattice indices. We frame the output of both models as coordinates by using that Owen \emph{et al.} \cite{Owen2020} assume their xanthophore and iridophore grid squares are $20$ $\mu$m wide, and their $M$ grid squares are $40$ $\mu$m wide (e.g., we map the indices $(i,j)$ of a melanophore to $(x,y)$-coordinates $(40i, 40j)$). At this stage, all of the ABM \cite{volkening2018} patterns are on domains of size $3.71$ mm long and $2.215$~mm high. The domains associated with CA \cite{Owen2020} patterns are all within one grid step of $6.93$ mm in length, and they vary from $1.56$ to $3.08$ mm in height (in our $1000$ simulations), since growth is uncoupled in the horizontal and vertical directions. To account for the difference in domain length between the models, we crop CA patterns, only considering the cells in a region of length $3.71$~mm in the center of the domain. Following the approach in \cite{McGuirl2020}, we remove the cells that fall within the top $10$\% or bottom $10$\% of the domain height for both the ABM and CA data, since stripes are often messy and not fully formed in those regions.

We crop patterns vertically using a percentage---rather than a strict number---based on our observations that both models have been developed to produce the same number of stripes and interstripes at a standard length of $12.63$ mm. In the ABM \cite{volkening2018}, the result is patterns on domains of cropped height $1.772$ mm. CA \cite{Owen2020} cells fall within domains of height between $1.248$ and $2.464$ mm, depending on the simulation, after our cropping step. (When CA domains are taller, stripes and interstripes are also wider.) We show example patterns highlighting the cells that have been cropped out from the top and bottom of the domains in Figure~\ref{fig:outliers}.  At the end of Step $1$, we have formatted the output of both models as coordinates of cell positions in micrometers on domains of length $3.71$ mm and variable height.

\begin{figure}[t!]
\centering
\includegraphics[width=0.9\textwidth]{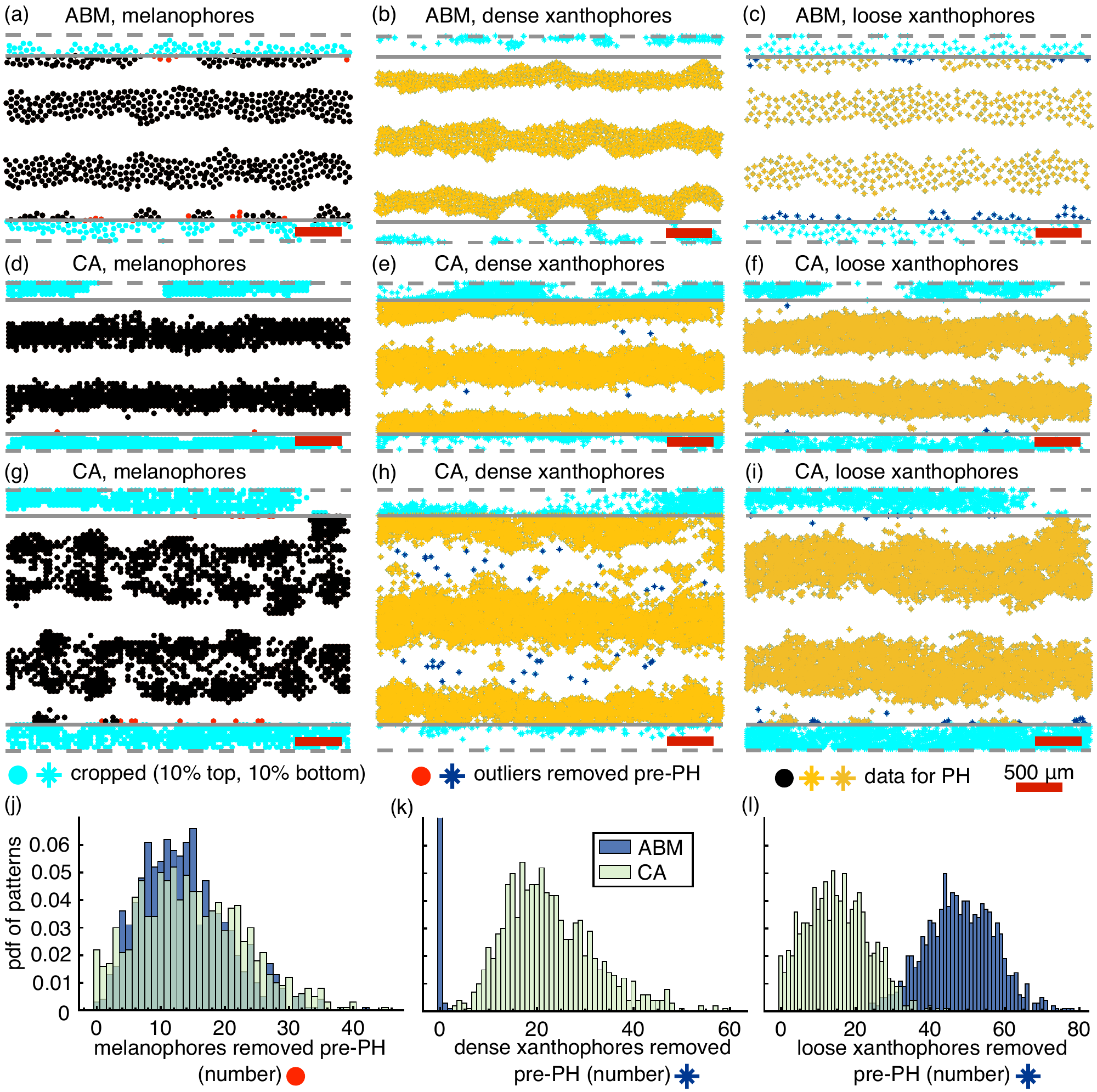}
\caption{Steps~$1$ and $2$: cropping and cleaning to remove outliers. For an example ABM \cite{volkening2018} pattern in (a) and two example CA \cite{Owen2020} patterns in (d, g), 
we indicate the $M$ cells that we crop out in Step $1$ (cyan) and remove as outliers in Step $2$ (red).  For the same simulations, the (b, e, h) $X^\text{d}$ and (c, f, i) $X^\ell$ cells that we crop out are cyan and remove as outliers are dark blue. The gray dashed bars are the domain boundaries, and the solid gray bars mark the cropped boundaries. Across $1000$ ABM and CA simulations, we show the distributions for the number of (j) $M$, (k) $X^\text{d}$, and (l) $X^\ell$ cells that we identify as outliers and remove in Step~$2$.
\label{fig:outliers}}
\end{figure}

\subsection{Step 2: Clean data to remove outlier cells} \label{sec:step2}

Our observations suggest that stripe and interstripe cells occupy distinct regions in ABM \cite{volkening2018} patterns. In comparison, the CA \cite{Owen2020} frequently produces outlier cells, such as an isolated black cell in the center of a gold interstripe; see Figure~\ref{fig:models}. Because the way that we compute persistent homology in Step $3$ is sensitive to outliers like these \cite{Fasy2014,Bendich2011}, we perform an initial cleaning step to remove stray cells. Focusing on the three cell types---$M$, $X^\text{d}$, and $X^\ell$---for which we compute persistent homology, we identify outliers by first finding the $N_{nn}=10$ nearest neighbors of each cell under the Euclidean norm on a domain that is periodic in $x$. We only consider neighbors of the same cell type, and we exclude each cell from counting itself. Using the distances $\{d_i\}_{i=1,...,N_{nn}}$ from the $j$th cell of type $c \in \{M, X^\text{d}, X^\ell\}$ to its nearest neighbors of type $c$, we compute $D^c_j = \sqrt{\sum_{i=1}^{N_{nn}} d_i^2/N_{nn}}$. If $D^c_j \ge \theta^c_m$, where $m \in \{\text{ABM}, \text{CA}\}$, then we classify the $j$th cell of type $c$ as an outlier and remove it. This captures that outliers are in low-density areas, so their distance from their ten nearest neighbors is comparatively large.

We use the thresholds $\theta^M_\text{ABM} = \theta^M_\text{CA}  = 220$ $\mu$m to identify melanophore outliers; $\theta^{X^\text{d}}_\text{ABM} = \theta^{X^\text{d}}_\text{CA} = 130$~$\mu$m to highlight dense xanthophore outliers; and $\theta^{X^\ell}_\text{ABM} = 215$ $\mu$m and $\theta^{X^\ell}_\text{CA} = 170$~$\mu$m to indicate loose xanthophore outliers. These initial cleaning thresholds, as well as our choice to take into account a cell's $N_{nn}=10$ nearest neighbors, are hyper-parameters that we do not investigate further. We chose them by observation, selecting values that identified outliers and fell in a range insensitive to changes. The larger distance thresholds for melanophores reflects that the distance between these cells is larger; see Figure~\ref{fig:models}(j). We show the distributions for the number of cells removed by type and model in Figures~\ref{fig:outliers}(j)--(l). The number of $M$ cells cleaned is comparable: a mean of $13.6$ cells with a standard deviation of $6.9$ cells for the ABM, and a mean of $14.8$ cells with a standard deviation of $8.3$ for the CA, across our $1000$ simulations. In comparison, the numbers of $X^\text{d}$ and $X^\ell$ cells removed differ between the models. On average, our methods identify zero $X^\text{d}$ outliers in the ABM \cite{volkening2018}, and $22.6$ $X^\text{d}$ outliers with a standard deviation of $9.2$ in the CA \cite{Owen2020}. For $X^\ell$ cells, we find a mean and standard deviation of $48.1 \pm 9.4$ outliers in the ABM and $15.2 \pm 8.4$ outliers in the CA.

The places where cells are being identified as outliers in the two models are also different; see Figure~\ref{fig:outliers}. As a result, cleaning the ABM data does not affect pattern quantification, and we only do this to apply the same set of steps to both models: McGuirl \emph{et al}. \cite{McGuirl2020} showed that persistent homology can be applied to quantify the ABM \cite{volkening2018} without any cleaning. This is because the ABM does not produce interstripe cells in stripe regions and vice versa; instead, as we show in Figures~\ref{fig:outliers}(a)--(c), the ``outliers" removed from ABM patterns tend to be cells at the upper and lower domain boundaries. In comparison, cleaning the CA data allows us to remove stray cells that occur between stripes and interstripes; see Figures~\ref{fig:outliers}(d)--(i). Without this cleaning, the presence of outlier cells in CA patterns makes interpreting topological summaries difficult in Step $3$ (\S \ref{sec:step3}). It is also worth noting that our initial cleaning does not account for the presence of $X^\text{d}$ spots (see, for example, Figure~\ref{fig:outliers}(h)) in some CA stripes; these spots have high enough local density that they are not identified as outliers, and we return to handling these pattern features in Step $4$  (\S \ref{sec:step4}).

\subsection{Step 3: Compute persistent homology to define stripe number and width}\label{sec:step3}

By building on the approach introduced by McGuirl \emph{et al.} \cite{McGuirl2020} (see \S \ref{sec:tda}), we use dimension-$1$ topological features, which correspond to loops or holes in our data, to count stripes and interstripes, define stripe and interstripe width, and flag broken or incomplete stripes and interstripes. 
Using the cropped, cleaned data from Step $2$, we calculate pairwise distances between $M$ cells, between $X^\text{d}$ cells, and between $X^\ell$ cells on a domain that is periodic in $x$.  (Because we crop the CA patterns so that our CA and ABM domains have the same length in Step $1$, we note that specifying boundary conditions that are periodic in $x$ on these cropped domains is an approximation for CA patterns.) We compute persistent homology in dimension $0$ and dimension $1$ based on these pairwise distances using Ripser \cite{Ripser}, enforcing a maximum distance of $650$~$\mu$m in our filtration. 

Specifically, we identify unbroken stripes and interstripes using the dimension-$1$ topological features (loops) for $X^\ell$ and $X^\text{d}$ cells, respectively, as below:
\begin{align}\text{number of unbroken stripes} &= \sum_\text{$X^\ell$ loops $i$} \textbf{1}_{b_i < B_m \Delta^m_{X^\ell X^\ell}}(i) \cdot \textbf{1}_{d_i - b_i > P_m}(i) \label{eq:stripes}\\
\text{number of unbroken interstripes} &= \sum_\text{$X^\text{d}$ loops $i$} \textbf{1}_{b_i < B_m \Delta^m_{X^\text{d}X^\text{d}}}(i) \cdot \textbf{1}_{d_i - b_i > P_m}(i),\label{eq:interstripes}
\end{align}
where $(b_i, d_i)$ are the $($birth, death$)$-coordinates of the $i$th loop; $d_i - b_i$ is the persistence of this feature; $\Delta^m_{cc}$ is the average distance between neighboring cells of type $c$ for the model $m \in \{\text{ABM}, \text{CA}\}$ across all $1000$ simulations (see Figure~\ref{fig:models}(j)); and the indicator function $\textbf{1}_\text{condition}(i) = 1$ if the given condition is met for topological feature $i$, and $0$ otherwise. The birth-time cutoff factor $B_m$ and persistence threshold $P_m$ with $m \in \{\text{ABM}, \text{CA}\}$ are our hyper-parameters for selecting the loops corresponding to stripes or interstripes in our data. In comparison, McGuirl \emph{et al.} \cite{McGuirl2020} used a strict birth-time threshold $B$ in place of our more flexible cutoff $B_m \Delta_{cc}$ that depends on the mean distance between cells. We sweep across these hyper-parameters in \S \ref{sec:results1} to identify choices that are robust for our focal models.

Following the approach \cite{McGuirl2020}, we use $M$ cells, in addition to $X^\text{d}$ and $X^\ell$ cells, when flagging patterns with stripe or interstripe breaks. Because we expect three interstripes in simulated zebrafish at the stage that we consider, we count a pattern as having broken interstripes if its number of interstripes in Eqn.~\eqref{eq:interstripes} is less than three. We classify a pattern as having broken stripes if two conditions are met: (1) its number of stripes in Eqn.~\eqref{eq:stripes} is less than two (the expected number of stripes for our simulated fish) and (2) its number of unbroken stripes from the perspective of melanophores is also less than two, where we define:
\begin{align}
\text{number of stripes from $M$ perspective} &= \sum_\text{$M$ loops $i$} \textbf{1}_{b_i < B_m \Delta^m_{MM}}(i) \cdot \textbf{1}_{d_i - b_i > P_m}(i)\label{eq:stripesM}.
\end{align}
McGuirl \emph{et al.} \cite{McGuirl2020} used both $M$ and $X^\ell$ cells to identify broken stripes in the ABM \cite{volkening2018} because Volkening \emph{et al.}'s model produces $X^\ell$ cells at low density; see Figure~\ref{fig:models}(d). This rationale applies in the reverse to the CA \cite{Owen2020}: $X^\ell$ cells appear at high density in interstripes, whereas $M$ cells are at comparatively low density, particularly in wide stripes; see Figures~\ref{fig:models}(h) and \ref{fig:outliers}(g). By flagging patterns as broken stripes only when the number of persistent loops is less than our expected number of stripes for both $M$ and $X^\ell$ cells, we allow for local dips in density for some stripe cells, as long as another type of stripe cell remains at consistently high enough density. We choose to focus on xanthophores and melanophores throughout our stripe-quantification pipeline; in the future, it would also be interesting to include iridophores.

In addition to counting (inter)stripes, we use persistent homology to compute stripe and interstripe width. Broadly, the death times of topological loops associated with interstripe cells provide information on stripe width, and vice versa, as below:
\begin{align} \text{stripe width} &= \underbrace{\text{med}_\text{$X^\text{d}$ interstripe loops $i$}d_i}_\text{median death time for interstripe loops} ~- \underbrace{\Delta_{MX^\text{d}}}_\text{boundary separation}\\
\text{interstripe width} &= \underbrace{\text{med}_\text{$X^\ell$ stripe loops $i$}d_i}_\text{median death time for stripe loops} ~- \underbrace{\Delta_{MX^\text{d}}}_\text{boundary separation},
\end{align}
where we define $X^\text{d}$ and $X^\ell$ dimension-$1$ topological features as ``interstripe loops" or ``stripe loops" using Eqns.~\eqref{eq:interstripes} and \eqref{eq:stripes}, respectively, and exclude features with infinite death times~$d_i$. Here $\Delta_{MX^\text{d}}$ is the mean distance between nearest melanophores and dense xanthophores at the stripe--interstripe interface for the specific simulation in question; see Figure~\ref{fig:models}(j). Subtracting $\Delta_{MX^\text{d}}$ prevents us from counting the interface-boundary distance twice. Our measurements of the width of any given (inter)stripe should be interpreted as maximum width\footnote{In comparison, McGuirl \emph{et al.} \cite{McGuirl2020} defined stripe width using the persistence of $X^\text{d}$ loops. This approach \cite{McGuirl2020} is similar to ours but does not account for the fact that the distance between $M$ and $X^\text{d}$ cells is larger than the distance between like cells. Additionally, due to an error in ball radius versus diameter, the summary statistics in \cite{McGuirl2020} report double (inter)stripe width.}. As an illustrative example of this, in Figure~\ref{fig:TDA}(a), the width of the dark stripe between the bottom two interstripes is slightly larger than $460$ $\mu$m $- \Delta_{MX^\text{d}}$, since a small amount of white space (a hole) is still visible in between the interstripes at $\varepsilon = 460$ $\mu$m.

\subsection{Step 4: Cluster cells into interstripes and further clean data}\label{sec:step4}
Broadly, we quantify curviness by clustering $X^\text{d}$ cells into interstripes in Step $4$ and then tracing out the upper and lower boundaries of each interstripe in Step~$5$. We use the method introduced in \cite{McGuirl2020} to specify the number of interstripe features (i.e., the target number of clusters $N_\text{cl}$) and apply single-linkage hierarchical clustering to the positions of $X^\text{d}$ cells to produce $N_\text{cl}$ clusters. In the ABM \cite{volkening2018}, these clusters generally correspond directly to interstripes. Because of differences in the noise and messiness present in ABM \cite{volkening2018} and CA \cite{Owen2020} patterns, however, we find that the same process does not transfer to the on-lattice model \cite{Owen2020}. Instead, as in the CA pattern in Figure~\ref{fig:outliers}(h), gold spots or stray $X^\text{d}$ cells that remain after our initial cleaning in Step $2$ can disrupt clustering. To account for these challenges, we introduce an iterative process to detect and remove clusters of $X^\text{d}$ cells that do not correspond to interstripes. This is a second round of data cleaning, removing more cells beyond those that we already removed in Step $2$.

Before identifying clusters corresponding to interstripes, the first step is determining the number of interstripe features. To account for gold bridges connecting broken stripes, McGuirl \emph{et al}'s method \cite {McGuirl2020} clusters $X^\text{d}$ cells into $N_\text{cl} = 3$ groups when a pattern has no breaks in black stripes; $N_\text{cl}= 2$~groups when the pattern has one broken stripe; and $N_\text{cl}= 1$ group otherwise. Our observations suggest that the models \cite{volkening2018,Owen2020} are meant to produce patterns with two stripes and three interstripes at the stage that we quantify. Similar to \cite{McGuirl2020}, we thus consider a pattern to contain one broken stripe when the number of stripes (from the perspective of $X^\ell$ or $M$ cells---or from the perspective of both) is one; see Eqns.~\eqref{eq:stripes} and \eqref{eq:stripesM}. As a book-keeping step, we also sort our matrix of cell coordinates by their $x$-coordinates before performing any clustering. Having cell coordinates in order based on their position from the left to right boundary of the domain is convenient later in Step $5$ (\S \ref{sec:step5}).

In an iterative process, we cluster our sorted $X^\text{d}$ cells into $N_\text{cl}$ groups, applying single-linkage hierarchical clustering to cell coordinates. (We use the Euclidean distance on a domain that is periodic in $x$.) We next calculate the number of cells in each cluster, and, if the number of cells is strictly less than our minimum cluster size $\tau^\text{min}_m$ with $m\in \{\text{ABM}, \text{CA} \}$, we remove all of the cells in that cluster from the domain. We then re-cluster the remaining $X^\text{d}$ cells into $N_\text{cl}$ groups, count the number of cells in our clusters, remove small clusters, and loop through again as needed. At the end of this process, the result is $N_\text{cl}$ clusters of $X^\text{d}$ cells that we interpret as corresponding to interstripes. We investigate the role of the hyper-parameter $\tau^\text{min}_m$ in \S \ref{sec:results1}.

\subsection{Step 5: Construct $\alpha$-shapes to quantify interstripe curviness}\label{sec:step5}

To compute the curviness of stripe--interstripe interfaces, the first step is extracting the boundaries of our clusters of $X^\text{d}$ cells from Step $4$. There are many ways to do this: for example, focusing on one interstripe at a time, McGuirl \emph{et al}. \cite{McGuirl2020} discretized space and then found the cells with the maximum and minimum $y$-coordinates in each grid step. Taken across all the grid positions, these highest and lowest cells trace out the upper and lower boundaries of an interstripe \cite{McGuirl2020}. Owen \emph{et al.} \cite{Owen2020} developed an algorithm for extracting the upper and lower boundaries of the center interstripe in the domain. This approach \cite{Owen2020} involves expressing the lattice of $X^\text{d}$ cells as a binary image and applying the built-in MATLAB function $\emph{bwmorph}$ to clean the data and bridge small gaps in interstripes. Owen \emph{et al}.\ then identify the height of the pixels associated with the upper and lower boundaries of the center interstripe; their methods \cite{Owen2020} focus on interstripe curviness as a whole---rather than stripe--interstripe boundary curviness---so they later take the  mean of these two curves and apply a smoothing algorithm. Importantly, both approaches \cite{McGuirl2020,Owen2020} rely on discretizing space: the lattice structure of the CA \cite{Owen2020} lends itself to image-processing techniques for binary images, and McGuirl \emph{et al}. \cite{McGuirl2020} introduce a spatial discretization to extract interstripe boundaries in the ABM.

We adapt $\alpha$-shapes \cite{Esurvey,Edelsbrunner1983} to develop a more flexible method for quantifying stripe--interstripe curviness in zebrafish patterns. As we discussed in \S \ref{sec:alpha}, $\alpha$-shapes provide a means of describing the boundary of a point cloud and are a generalization of its convex hull; see Figure~\ref{fig:alphaShape}. Here our point cloud consists of the coordinates of $X^\text{d}$ cells associated with each interstripe cluster from Step $4$. One interstripe at a time, we construct an $\alpha$-shape and extract the coordinates of the $X^\text{d}$ cells lying on its boundary using the built-in MATLAB functions \emph{alphaShape} and \emph{boundaryFacets} (with holes suppressed). This leads to a sequence of coordinates $(\textbf{X}^\text{d,bnd}_i)_{i=1,...,N^\text{bnd}}$ that are the vertices of our $\alpha$-shape, wrapping around the interstripe. We next find the first index $k$ in this sequence with $x$-coordinates equal to the maximum $x$-coordinates in the sequence. Because we sort the list of $X^\text{d}$ cells with respect to their $x$-coordinates in Step $4$ before clustering into interstripes, this index splits the boundary $X^\text{d}$ cells in a way that captures the top and bottom boundaries of the interstripe: specifically, we consider the sequences $(\textbf{X}^\text{d,bnd}_i)_{i=1,...,k}$ and $(\textbf{X}^\text{d,bnd}_i)_{i=k,...,N^\text{bnd}}$. To complete the process, we remove the cells in both of these sequences that fall within $100$~$\mu$m (inclusive) of the right or left boundaries of the domain and renumber indices as needed\footnote{Using $100$ $\mu$m here is a choice that we do not investigate further. Excluding this small region near the right and left boundaries of the pattern is fairer to both models. In particular, because we crop CA domains lengthwise in Step $1$, they could potentially be at a disadvantage when computing curviness, since ABM domains have the benefit of unaltered periodic boundary conditions in $x$. In addition to helping us separate cell coordinates into upper and lower boundaries, cropping the region of the pattern for which we compute interstripe curviness by $200$ $\mu$m puts both models on equal footing.}. This last step ensures that we have excluded the right and left boundaries of our $X^\text{d}$ point cloud. The result is two sequences of cell coordinates tracing out the top and bottom boundaries of the interstripe in question.

For each sequence of cell coordinates $(\textbf{X}_i^\text{d,bnd})_{i=1,...,N}$ representing the boundary of an interstripe, we compute the length of the associated piecewise linear curve. We then use the length of a perfectly straight interstripe to compute curviness, similar to \cite{McGuirl2020}, as below:
\begin{align} \text{curviness of boundary} &= 100 \times \left(\frac{\sum_{i=1}^{N-1}||\textbf{X}^\text{d,bnd}_{i} - \textbf{X}^\text{d,bnd}_{i+1}||}{\max_i (x_i^\text{d,bnd}) - \min_i (x_i^\text{d,bnd})} -1 \right)\label{eq:curve}
\end{align}
where $||\textbf{x}|| = \sqrt{x_1^2 + x_2^2}$ and $\textbf{X}^\text{d,bnd}_i = (x^\text{d,bnd}_i,y^\text{d,bnd}_i)$. Repeating this process for each of the $N_\text{cl}$ interstripe clusters from Step $4$, we generate $2\times N_\text{cl}$ measurements of boundary curviness. Across all of these boundaries, we exclude the curves with the maximum and minimum $y$-coordinates, since we find that interstripes in some CA patterns (e.g., see Figure~\ref{fig:outliers}) extend beyond the limits of our focal domain. As a result, these uppermost and lowermost curves are often simply tracing the upper and lower boundaries of the rectangular domain. For the remaining $2N_\text{cl}-2$ boundary curves, we take the mean of their curviness values from Eqn.~\eqref{eq:curve} and define this as our measurement of pattern curviness.

Our method for going from $\alpha$-shape vertices to two sequences of cells representing the upper and lower boundaries of an interstripe does not work in all cases. In particular, when a pattern is assigned $N_\text{cl} < 3$ from Step $4$ this means that at least one ``interstripe cluster" is, in fact, two interstripes connected by a bridge across a broken black stripe. Our approach does not extract boundaries that make intuitive sense in this case. Thus, in our results in \S \ref{sec:results}, we present summary statistics on interstripe--stripe boundary curviness for unbroken patterns. It would be interesting in the future to build on our methods to define and extract the upper and lower boundaries of interstripes when interruptions are present in the pattern. We sugggest that the single and visible hyper-parameter $\alpha$ associated with constructing $\alpha$-shapes makes them a nice tool to use when quantifying curviness, and we sweep across different values of $\alpha_m$ for $m \in \{\text{ABM},\text{CA}\}$ to understand its impact in \S \ref{sec:results1}.

\section{Results: Quantitative study of on- and off-lattice microscopic models}\label{sec:results}

We now apply our methods from \S \ref{sec:methods} to quantify stripe patterns resulting from two biologically detailed models of zebrafish, the ABM \cite{volkening2018} and CA \cite{Owen2020}. We begin by sweeping through different choices of the hyper-parameters in Figure~\ref{fig:pipeline} to determine their effects on pattern quantification (\S \ref{sec:results1}). We then select values for our hyper-parameters that allow us to put simulated patterns from the models \cite{volkening2018,Owen2020} on equal footing and present a large-scale, discerning study of these models in \S \ref{sec:results2}. See the caption of Figure~\ref{fig:pipeline} for a summary of the baseline hyper-parameter values that we establish in \S \ref{sec:results1} and use in \S \ref{sec:results2}.

\subsection{Robustness study: Choosing hyper-parameters}\label{sec:results1}

Our methods in Figure~\ref{fig:pipeline} depend on the choice of values for eight main hyper-parameters: four hyper-parameters associated with our initial data cleaning in Step $2$, two hyper-parameters that emerge in Step~$3$ when we interpret persistent homology, one hyper-parameter related to clustering cells into interstripes in Step~$4$, and a final hyper-parameter that controls how tightly we estimate interstripe boundaries in Step~$5$. Because it is not just choices of computational implementation, but also choices in methods for quantification, that affect how we interpret the output of different models, here we sweep across values of our hyper-parameters to better understand their roles and select values that are biologically meaningful and robust. We focus on the four hyper-parameters involved in implementing and interpreting persistent homology and constructing $\alpha$-shapes in Steps~$3$--$5$. (See \S \ref{sec:conclusions} for a discussion of the other hyper-parameters and additional choices in our pipeline.)

\begin{figure}[t!]
\centering
\includegraphics[width=0.9\textwidth]{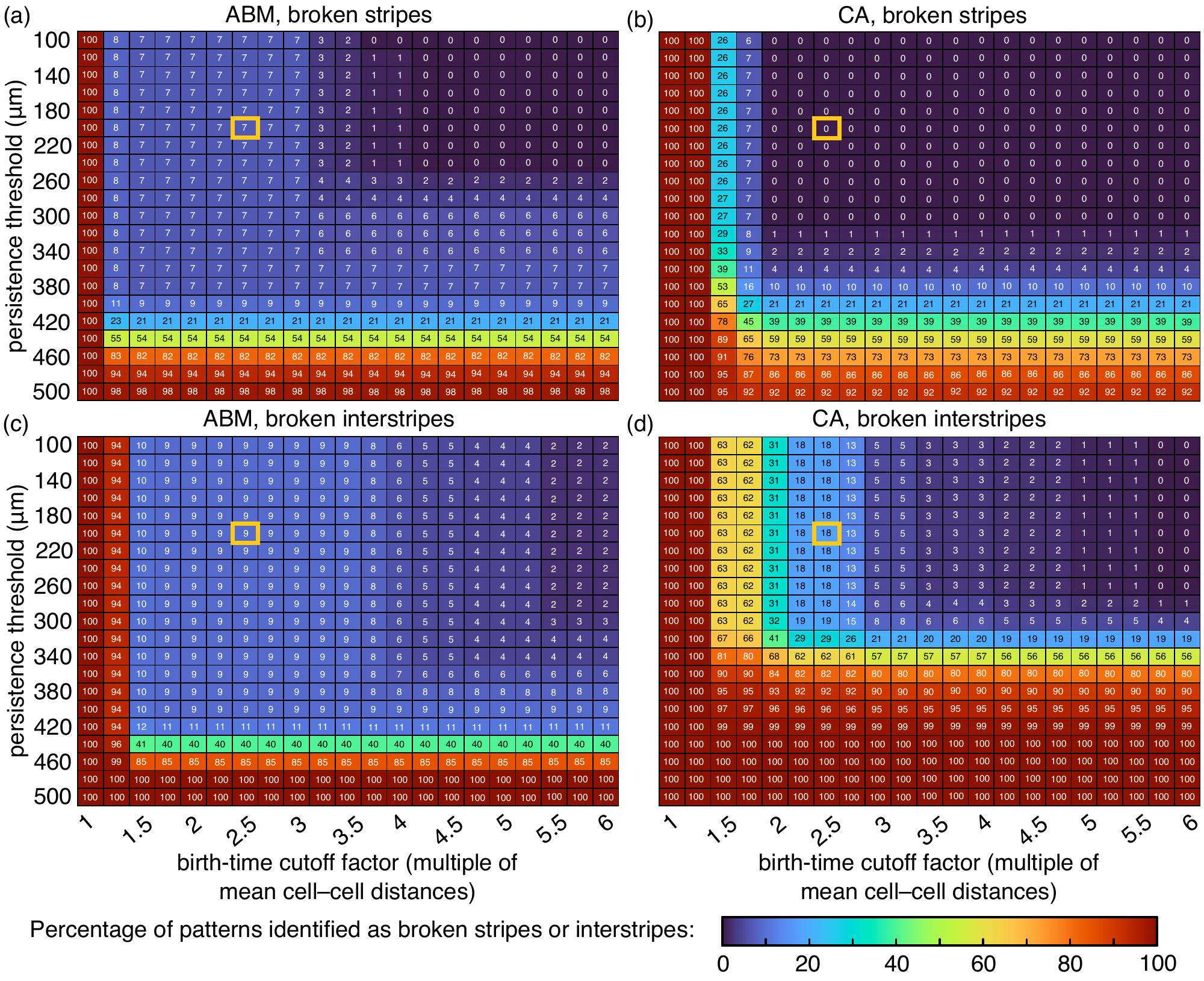}
\caption{Effects of the birth-time cutoff factor $B_m$ and persistence threshold $P_m$ on identifying broken patterns in Step $3$ (Eqns.\ \eqref{eq:stripes}--\eqref{eq:stripesM}). Results are based on $100$ ABM \cite{volkening2018} and $100$ CA \cite{Owen2020} patterns, cropped and cleaned as described in \S \ref{sec:step1} and \S \ref{sec:step2}, that we quantify under each choice of $(B_m,P_m)$ given. Here $B_m$ is a factor multiplied by the appropriate mean cell--cell distance (see Figure~\ref{fig:models}(j)). (a) The number of ABM patterns flagged for broken stripes decreases as  we relax how early loops must be born to count as stripes. (b) Our results for CA patterns show similar overall dependence on $B_m$ and $P_m$, and we note that broken stripes are rare in CA patterns. (c) For large values of $P_m$, there are not sufficient topological loops of $X^\text{d}$ cells with persistence $P_m$, so the number of ABM patterns flagged as having broken interstripes is $100$\%. (d) Our observations suggest that CA patterns feature some $X^\text{d}$ cells that ``infringe" on black stripe territory; this may be visible in $100$\% of CA patterns being flagged for broken interstripes at a lower value of $P_m$ relative to ABM patterns. We highlight the baseline hyper-parameter values that we use in \S \ref{sec:results2} in yellow. The hyper-parameters $B_m$ and $P_m$ also play a role in how we quantify (inter)stripe width; see Appendix \ref{sec:supp}. \label{fig:sweeping1}}
\end{figure}

To interpret birth and death times of dimension-$1$ topological features as information about (inter)stripe width and number in Step~$3$ \cite{McGuirl2020}, we rely on a birth-time cutoff factor $B_m$ and persistence threshold $P_m$, where $m \in \{\text{ABM},\text{CA}\}$ denotes the potential for different values for the two models \cite{volkening2018,Owen2020}. For values $B_m = 1, 1.25, 1.5, ..., 5.5,5.75, 6$ and $P_m =100,120,140,...,460,480,500$ $\mu$m in Eqns.\ \eqref{eq:stripes}--\eqref{eq:stripesM}, we quantify $100$ ABM and CA patterns, cropped and cleaned as described in \S \ref{sec:step1} and \S \ref{sec:step2}, under different choices of $B_m$ and $P_m$. In Figure~\ref{fig:sweeping1}, we show the effects of these hyper-parameters on the percentage of patterns that our methods identify as having broken stripes or interstripes. (We find that stripe and interstripe width is less sensitive to $B_m$ and $P_m$; see Appendix \ref{sec:supp}.) The extreme choice of $B_\text{m}=1$ requires that loops corresponding to (inter)stripes be born at a scale strictly less than the mean distance between cells, and this intuitively leads to no patterns being identified as having three interstripes and two stripes; as expected, our methods flag $100$\% of patterns as broken when $B_m$ is too low in Figure~\ref{fig:sweeping1}. Similarly, a persistence threshold $P_m$ that is too large---e.g., beyond the width of interstripes---means our methods flag all patterns as broken because the number of dimension-$1$ topological features with large enough persistence is below our target of three.

Beyond low $B_m$ and high $P_m$ values, which are restrictive constraints on our inequalities in Eqns.\ \eqref{eq:stripes}--\eqref{eq:stripesM}, we suggest that $B_m \ge 4$ is overly permissive of gaps between cells in (inter)stripes. This is based on empirical observations that zebrafish stripes are roughly $7$--$12$ cells wide \cite{Nakamasu}, so $B_m \ge 4$ means that stripes featuring gaps between adjacent cells comparable to half a stripe width pass through our methods without being flagged as broken. If we focus on $B_m < 4$ in Figure~\ref{fig:sweeping1}, there is a wide range of persistence thresholds under which our methods are insensitive. Choosing $P_\text{ABM} \in [100,340]$ and $P_\text{CA} \in [100,260]$ $\mu$m does not affect our results. (If we use different hyper-parameters for different cell types, we can consider a wider range for stripes, but we apply the same cutoff in Eqns.~\eqref{eq:stripes}--\eqref{eq:stripesM}.) In agreement with \cite{McGuirl2020}, we thus select $P_\text{ABM}=P_\text{CA} = 200$~$\mu$m as our baseline value. The sharp drop in ABM interstripes (Figure~\ref{fig:sweeping1}(c)) flagged as broken between $B_\text{ABM} = 1.25$ and $1.5$, followed by a fairly insensitive region, suggests that we should select $B_\text{ABM} \in [1.5,3.75]$. For CA patterns, the number of interstripes flagged as broken  (Figure~\ref{fig:sweeping1}(d)) repeatedly drops by about half as $B_\text{CA}$ increases from $1.25$ to $2.25$. In the end, choosing a value for $B_m$ is a question of how large of a gap between adjacent cells we permit before calling an (inter)stripe broken, and we set $B_\text{ABM} = B_\text{CA} =2.5$ cell--cell distances as our baseline value.

\begin{figure}[t!]
\centering
\includegraphics[width=0.9\textwidth]{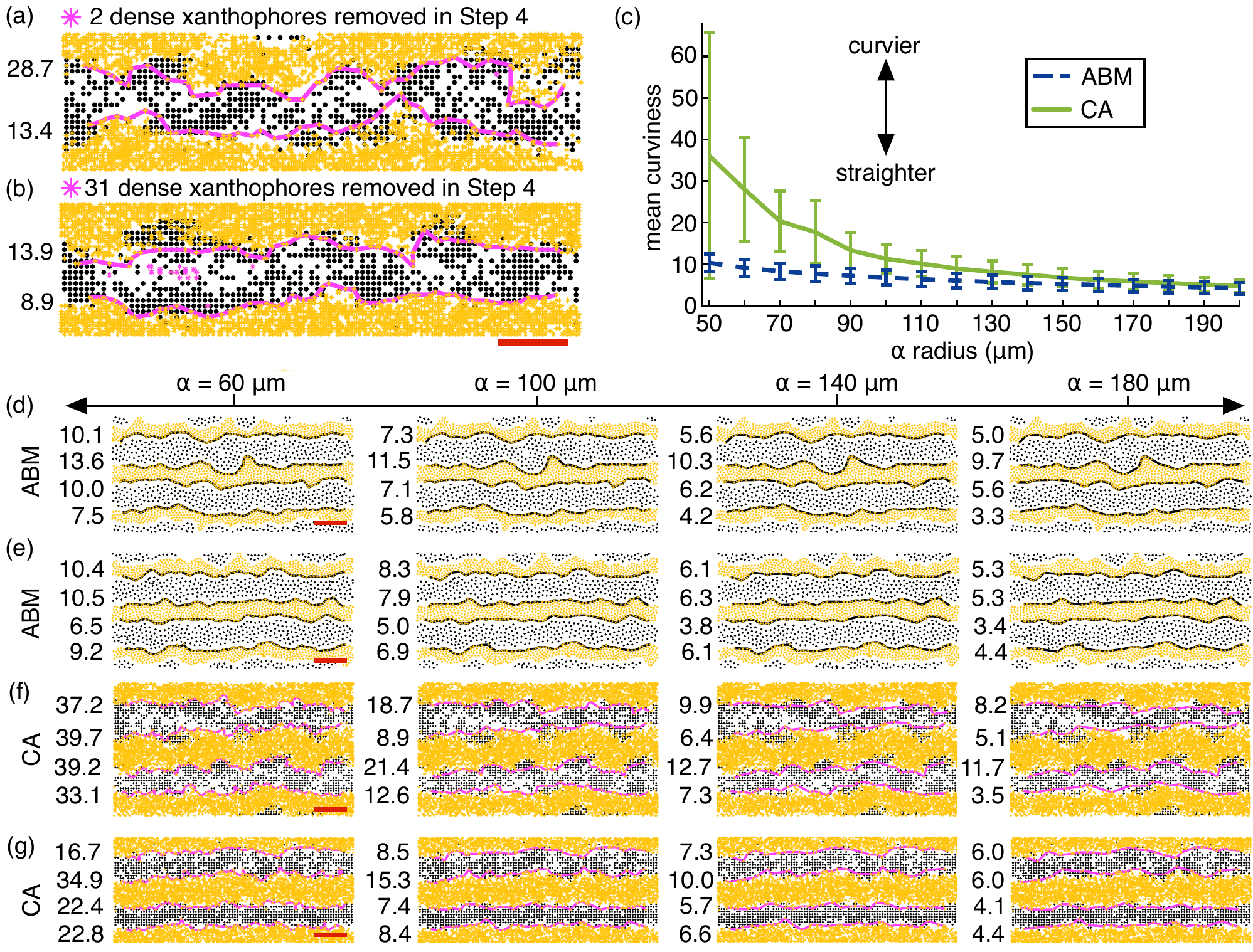}
\caption{Effects of the minimum cluster size $\tau^\text{min}_m$ and $\alpha$-radius $\alpha_m$ on stripe--interstripe curviness in Steps~$4$--$5$. Our results are for $100$ ABM \cite{volkening2018} and CA \cite{Owen2020} patterns, cropped and cleaned in Steps~$1$--$2$. We show summary statistics for the subset of these patterns classified as unbroken in Step $3$ with the baseline hyper-parameter values established in Figure~\ref{fig:sweeping1}. (a)--(b) Some CA patterns \cite{Owen2020} require further cleaning (i.e., removal of stray $X^\text{d}$ cells and spots) to help ensure the $N_\text{cl}$ groups that we cluster $X^\text{d}$ cells into in Step $4$ correspond to interstripes. We show a closer look at an example CA pattern with (a) two $X^\text{d}$ cells removed before computing curviness, and one with (a) a larger spot removed. Removed cells are shown in magenta. Once $\tau^\text{min}_\text{CA}$ is large enough to fall into the range $[20,100]$ cells, our methods are insensitive to this choice. Based on \cite{McGuirl2020}, $\tau^\text{min}_\text{ABM} = 0$ cells (i.e., no additional cleaning) is sufficient for ABM patterns. Our choice of baseline value $\tau^\text{min}_m = 60$ cells for $m\in \{\text{ABM}, \text{CA}\}$ means that $X^\text{d}$ clusters must contain at least $60$ cells to be considered interstripes, and smaller clusters are removed. (c) As the radius $\alpha_m$ used to compute $\alpha$-shapes grows, interstripe boundaries get closer to the convex hull of our $X^\text{d}$ data. (d)--(e) We show two example ABM patterns with interstripe boundaries (blue) extracted under different choices of $\alpha_\text{ABM}$ and (f)--(g) two example CA patterns with interstripe boundaries (magenta) associated with different $\alpha_\text{CA}$ values. The numbers to the left denote curviness; see Eqn.~\eqref{eq:curve}. We use $\alpha_m  = 100$ $\mu$m for $m \in \{\text{ABM},\text{CA}\}$ as our baseline value in \S \ref{sec:results2}.\label{fig:sweeping2}}
\end{figure}

After using our baseline values of $B_m$ and $P_m$ to output the number of stripes and interstripes in Step $3$, we turn to the hyper-parameters $\tau^\text{min}_m$ and $\alpha_m$ that affect our measurements of curviness. We evaluate $100$ CA and ABM patterns, excluding those with broken (inter)stripes, under minimum $X^\text{d}$ cluster size $\tau^\text{min}_m=20,30,...,90,100$ cells and $\alpha$ radius $\alpha_m=50,60,..,190,200$ $\mu$m. Because McGuirl \emph{et al}. \cite{McGuirl2020} clustered cells into interstripes for ABM \cite{volkening2018} patterns without any need to remove non-interstripe clusters first, we note that $\tau^\text{min}_\text{ABM}=0$ cells is valid. For CA patterns, as we show in Figures~\ref{fig:sweeping2}(a)--(b), boundary roughness makes it challenging to visually specify which $X^\text{d}$ cells should be removed before computing curviness. Sweeping across values of $\tau^\text{min}_\text{CA}$ and $\alpha_\text{CA}$, all values of the minimum cluster size $\tau^\text{min}_m$ between $20$ and $100$ cells lead to the same curviness results, so we do not show this heatmap. In contrast, choosing $\tau^\text{min}_\text{CA}$ much lower than $20$ cells does not reliably group $X^\text{d}$ cells into $N_\text{cl}$ clusters that are interstripes, and it causes errors in our algorithm: when $\tau^\text{min}_\text{CA}$ is too low, outlier cells in CA patterns are sometimes selected as one of the interstripe clusters in Step~$4$. We thus choose $\tau^\text{min}_\text{CA}= 60$ cells as our baseline value, and set $\tau^\text{min}_\text{ABM}=\tau^\text{min}_\text{CA}$ for consistency.

As we show in Figure~\ref{fig:sweeping2}(c), our results on stripe--interstripe boundary curviness for unbroken patterns across $100$ ABM and CA simulations depend strongly on the choice of radius $\alpha_m$ that we use when constructing $\alpha$-shapes in Step $5$. This is particularly the case for CA patterns, which feature jagged interstripe edges. Selecting $\alpha_m$ means balancing sharply tracing out interstripe boundaries with relaxing our curves toward the convex hull of $X^\text{d}$ positions. To gain intuition into the effects of $\alpha_m$, we show the interstripe boundaries that we extracted and associated boundary curviness under different choices of $\alpha_m$ for two example ABM patterns in Figures~\ref{fig:sweeping2}(d)--(e) and two example CA patterns in Figures~\ref{fig:sweeping2}(f)--(g). The hyper-parameter $\alpha_m$ essentially controls how ``forgiving" our methods are to peaks and crevices in $X^\text{d}$ organization.  Because the boundary curves for $\alpha_\text{CA} = 60$ $\mu$m are very rough and convoluted, we suggest that this value focuses on too fine of details in CA patterns. On the other hand, selecting $\alpha_\text{CA} = 140$~$\mu$m or $180$ $\mu$m Figures~\ref{fig:sweeping2}(f)--(g) leads to boundary curves that miss structure and undulations in $X^\text{d}$ organization that we suggest are meaningful. Supported by these observations and the fact that our curviness measurements for CA patterns show much less variance and change less rapidly as a function of $\alpha_\text{CA}$ after about $90$ $\mu$m in Figure~\ref{fig:sweeping2}(c), we select $\alpha_\text{CA} = 100$ $\mu$m as our baseline value. Again for consistency, we set $\alpha_\text{ABM} = 100$~$\mu$m.

\subsection{Quantitative description of microscopic models at large scale}\label{sec:results2}

With the baseline values of our hyper-parameters from \S \ref{sec:results1} (see the caption of Figure~\ref{fig:pipeline} for a summary), we now present a large-scale study of $1000$ ABM \cite{volkening2018} and $1000$ CA \cite{Owen2020} patterns. As we show in Figure~\ref{fig:baseline}, our methods provide information on (inter)stripe width, stripe--interstripe boundary curviness, and the fraction of patterns with overall imperfections (e.g., broken stripes). We also track the number of non-interstripe $X^\text{d}$ clusters removed in Step~$4$ before computing boundary curviness. Interpreting our results depends on viewpoint: from one perspective, the ABM and CA patterns are remarkably similar, particularly given the differences in noise structure and outlier cells that we accounted for in \S \ref{sec:step2}. From another perspective, the results in Figure~\ref{fig:baseline} show that the ABM and CA---despite being biologically similar---do produce different behaviors, most notably related to variability across simulations. Our quantitative approach allows us to move beyond viewing a few sample patterns from the stochastic models \cite{volkening2018,Owen2020} to better understand the features that these models give rise to at large scale.

\begin{figure}[t!]
\centering
\includegraphics[width=0.9\textwidth]{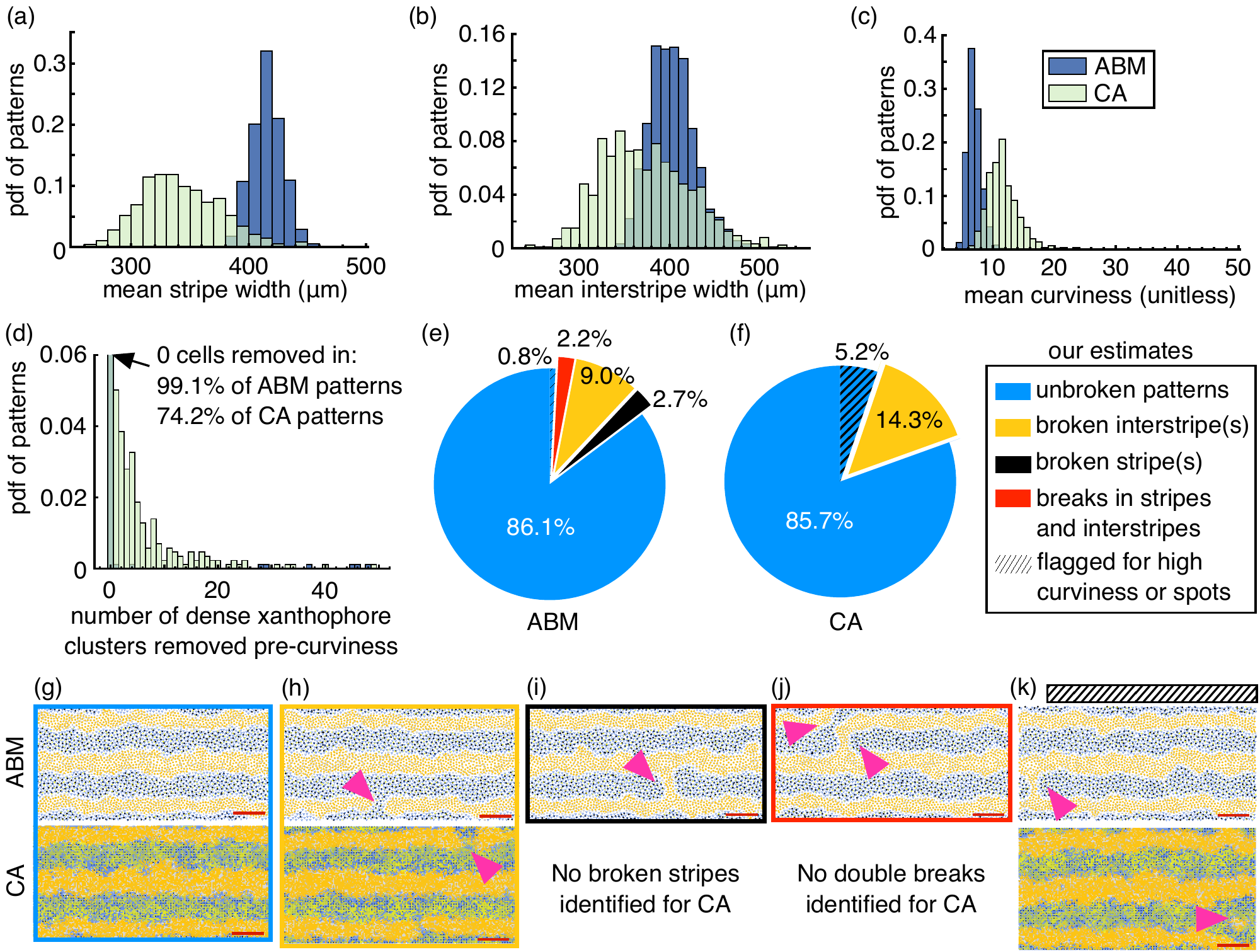}
\caption{Baseline quantification of $1000$ patterns generated from the ABM \cite{volkening2018} and CA \cite{Owen2020}. Distributions in (a)--(d) are for patterns identified as unbroken (namely, patterns with three interstripes and two stripes). (a) The overlap of stripe and interstripe cells in CA \cite{Owen2020} patterns may play a role in why our methods estimate stripe width as narrower in CA patterns than in ABM \cite{volkening2018} patterns. (b) With respect to interstripe width and other pattern features, CA \cite{Owen2020} patterns are more variable than ABM \cite{volkening2018} patterns. (c) High curviness may be associated with the presence of spots in patterns or with incorrectly classified broken patterns; see Figure~\ref{fig:outliers}. (d) To help account for spots in CA patterns, we further clean the data in Step~$4$; here we show the number of $X^\text{d}$ clusters removed before computing curviness. (e) ABM \cite{volkening2018} patterns are susceptible to breaks in stripes and interstripes, while (f) CA \cite{Owen2020} are most susceptible to breaks in interstripes. For the ABM (top row) and CA (bottom row), we show example patterns that our methods identify (g) as unbroken, (h) as having broken interstripes, (i) as having unbroken stripes, or (j) as having breaks in stripes and interstripes. (k) While our methods classify these patterns as unbroken, they feature imperfections and represent false negatives in the form of a narrow break (ABM) and prominent spot (CA). Red scale bar in (g)--(k) is $500$ $\mu$m.\label{fig:baseline}}
\end{figure}

As we show in Figure~\ref{fig:baseline}(a), the mean stripe width (across $1000$ simulations) for the ABM \cite{volkening2018} is about $416$ $\mu$m, and the mean stripe width for the CA \cite{Owen2020} is about $343$ $\mu$m, a difference that corresponds to less than two $M$ cells. The mean interstripe widths for the ABM and CA are closer: $403$~$\mu$m and $373$ $\mu$m, respectively (see Figure~\ref{fig:baseline}(b)). This is a difference of less than one $X^\text{d}$ cell from the perspective of the ABM \cite{volkening2018} and less than two $X^\text{d}$ cells from the perspective of the CA \cite{Owen2020}. In terms of stripe--interstripe boundary curviness based on Eqn.~\eqref{eq:curve} (Figure~\ref{fig:baseline}(c)), the mean curviness of ABM patterns not flagged as broken is about $7.0$ with a standard deviation of about $1.8$. Similarly, the mean curviness of CA patterns is about $11.8 \pm 3.0$. As we show in Figure~\ref{fig:baseline}(d), across the portion of our $1000$ ABM and CA simulations that are not flagged as broken, $99.1$\% of ABM patterns do not undergo additional cleaning of $X^\text{d}$ cells in Step $4$. In comparison about a quarter of unbroken CA patterns contain small $X^\text{d}$ clusters that we remove before computing interstripe--stripe boundary curviness.

\begin{figure}[t!]
\centering
\includegraphics[width=0.9\textwidth]{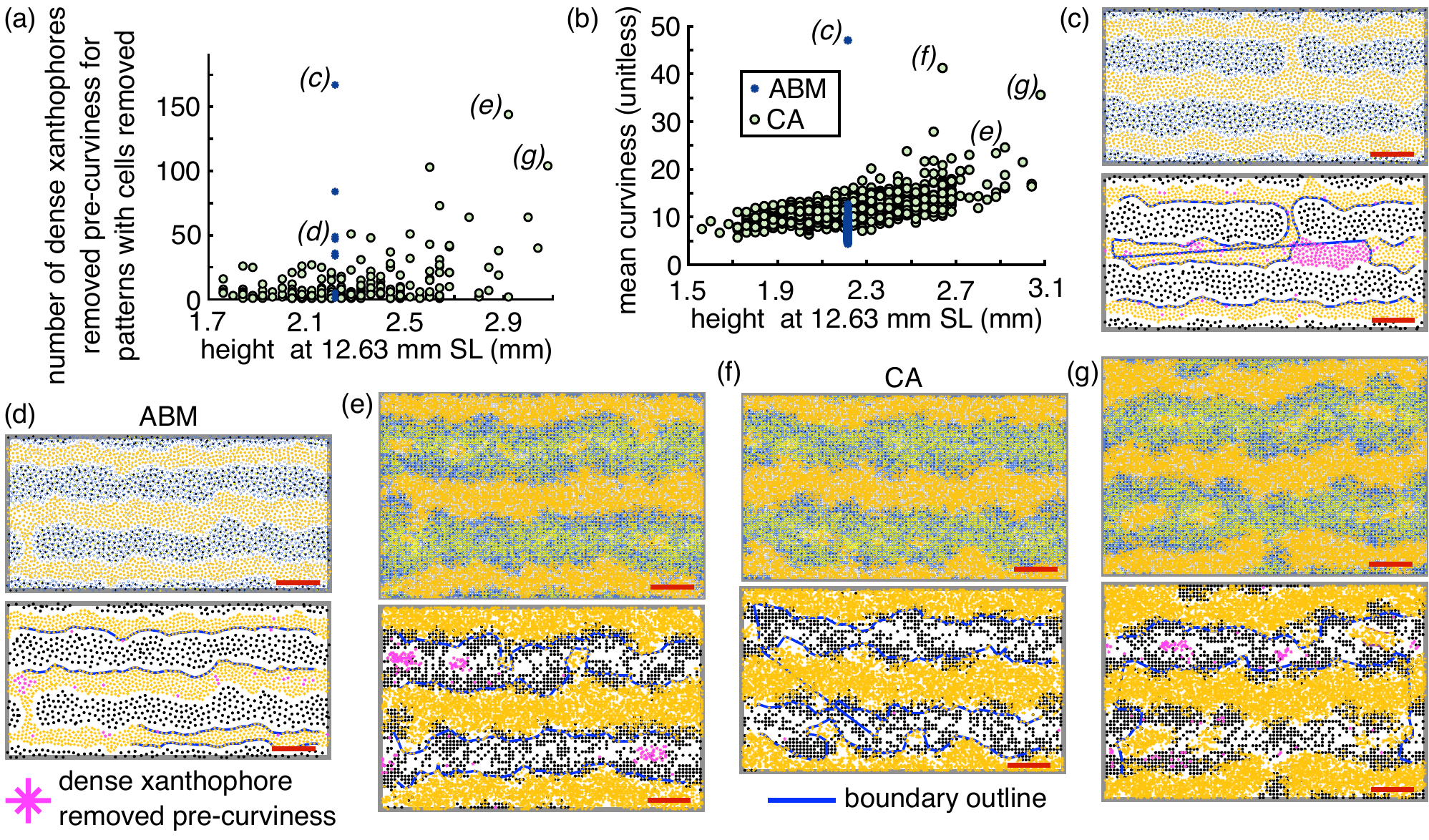}
\caption{False negatives, spots, and the role of domain growth. (a) Across the subset of $1000$ ABM and CA patterns identified by our methods as unbroken, we show the total number of $X^\text{d}$ cells removed in Step~$4$ versus the domain height. Because the ABM \cite{volkening2018} has deterministic domain growth, all scatter points are stacked at the same height; there are only eight patterns with more than zero $X^\text{d}$ cells cleaned in Step~$4$. The CA \cite{Owen2020} includes stochastic domain growth, meaning that domains of the same length have different heights. Interestingly, the spread of scatter points spreads with tall CA domains. (b) Removal of $X^\text{d}$ cells in Step~$4$ and our curviness computations in Step~$5$ are related, since interstripe boundaries must curve and bend to encompass any $X^\text{d}$ cells remaining in the pattern. We observe that the mean curviness of CA patterns increases and spreads with increasing domain height, particularly after CA domains become taller than ABM domains. The letters in (a, b) refer to examples in (c)--(g). (c, d) We show two example ``unbroken" ABM \cite{volkening2018} patterns; the top row provides the full pattern, and the bottom row focuses on $M$ and $X^\text{d}$ cells for clarity, with $X^\text{d}$ cells removed in Step~$4$ indicated in magenta and interstripe boundaries shown in blue. We see odd behavior because $X^\text{d}$ cells are being forced into three clusters, when only two are present. A narrow bridge is clearly visible in these patterns, and we conclude that all eight ABM \cite{volkening2018} patterns with any removal of $X^\text{d}$ cells in Step~$4$ are false negatives. (e)--(g) CA patterns with large numbers of $X^\text{d}$ cells removed in Step~$4$ and high curviness often feature $X^\text{d}$ spots. Comparing (f) and (g), for example, it can be challenging to judge what constitutes a spot versus a finger of an interstripe that should be included in its boundary. We flag CA \cite{Owen2020} with very high curviness or a large number of $X^\text{d}$ clusters removed in Step~$4$ as potentially spotted; see Figure~\ref{fig:baseline}. \label{fig:height}}
\end{figure}

At the pattern scale, we show the percentage of ABM and CA patterns that our methods identify as unbroken, as having broken interstripes only, as having broken stripes only, or as having breaks in stripes and interstripes in Figures~\ref{fig:baseline}(e)--(f). We find that our quantification pipeline reports a remarkably similar percentage of ABM \cite{volkening2018} and CA \cite{Owen2020} patterns as unbroken. Beyond this, ABM patterns \cite{volkening2018} display breaks of different types, whereas CA patterns \cite{volkening2020} are most susceptible to breaks in interstripes. Additionally, we flag---as a subset of patterns classified as ``unbroken"---a rough estimate of the ABM and CA patterns that we suggest are false negatives, meaning simulations that our methods should have classified as having pattern errors but did not.

How we flag patterns as potential false negatives is model specific. In particular, based on our observations and the results \cite{McGuirl2020}, we expect dependable separation of $X^\text{d}$ cells into interstripes in ABM \cite{volkening2018} patterns. Thus, we suspect that the eight out of $1000$ ABM patterns that are classified as unbroken, yet led to removal of some positive number of $X^\text{d}$ cells in Step~$4$, are, in fact, broken. As we show in Figures~\ref{fig:height}(c)--(d) for two examples among these eight patterns, this is indeed the case. The eight ABM false negatives occur because $B_\text{ABM}$ is just large enough so that the number of stripes from either the perspective of $X^\ell$ or $M$ cells in Eqns.~\eqref{eq:stripes} or \eqref{eq:stripesM} is two, allowing the narrow gaps in Figures~\ref{fig:height}(c)--(d) to pass by undetected. As a result, interstripes that are actually connected by a bridge are forced into three clusters, and the result is haphazard removal of $X^\text{d}$ cells in Step $4$.

On the other hand, we flag $52$ ``unbroken" CA patterns as potential false negatives in Figure~\ref{fig:baseline}(f). These fall into two groups: (1) patterns that require more than $25$ $X^\text{d}$ clusters to be removed in Step~$4$ before computing curviness, or (2) patterns with at least one stripe--interstripe boundary with curviness greater than $25$. Patterns meeting these conditions are rare, as indicated by the distributions in Figures~\ref{fig:baseline}(c)--(d). We show a few example CA patterns among these $52$ potential false negatives in Figures~\ref{fig:height}(e)--(g): these patterns clearly include large $X^\text{d}$ spots or tortuous boundaries. Because our methods do not directly quantify spots, flagging CA patterns in Figure~\ref{fig:baseline}(f) with very high curviness or many outlier $X^\text{d}$ cells serves as a rough proxy for indicating that a subset of ``unbroken" patterns feature spot imperfections. Adding on to our methods and those in \cite{McGuirl2020} to identify and characterize spots in CA patterns directly is an important direction for future research. 

To further check our methodology, we viewed $200$ patterns for each model. For the ABM \cite{volkening2018}, $171$ of these patterns ($85.5$\%) were classified as unbroken by our methods, and our qualitative observations agree in $170$ cases. The single false negative features a narrow gap in a black stripe, and, as discussed above, it is flagged as potentially broken because it leads to more than zero $X^\text{d}$ cells removed in Step $4$. We agree with the $18$ ($9$\%), $8$ ($4$\%), and $3$ ($1.5$\%) patterns identified by our methods as having breaks in interstripes, stripes, or both, respectively. We had more difficulty classifying the CA \cite{Owen2020} simulations visually, since these patterns feature more outliers and cells grouped less distinctly into stripes and interstripes. Across $200$ CA patterns, our methods identified $0$ with broken stripes (we agree) and $34$ ($17$\%) with broken interstripes (we suggest $7$ of these are false positives and not broken). Among the $166$ ($83$\%) patterns that our methods identified as unbroken, we suggest four of these have interstripe breaks and three feature one or more prominent spots. Ten more of these unbroken patterns were flagged for high curviness as discussed above, and they all have at least one large visual spot. Our visual checks add confidence that our methods are capturing a meaningful overall picture of the patterns generated by both models \cite{volkening2018,Owen2020}.

As a final note, we comment on a few interesting features of our curviness results in Figure~\ref{fig:baseline}(c). First, because $X^\text{d}$ cells are more dense in CA patterns than in ABM patterns (see Figure~\ref{fig:models}), we thought this might lead our methods to interpret CA patterns as curvier than ABM patterns. To better understand this, we calculated the mean distance between pairs of $X^\text{d}$ cells that share an edge in each graph representing an interstripe boundary, across all unbroken patterns. We found that the mean edge length in ABM boundary curves is about $62.7$ $\mu$m---a little under twice the mean distance between neighboring $X^\text{d}$ cells---and the mean edge length in CA boundary curves is about $83.8$ $\mu$m---just under four times the mean distance between nearest $X^\text{d}$ cells. This adds support to the conclusion that the difference in curviness between ABM \cite{volkening2018} and CA \cite{Owen2020} patterns is an inherent feature of the models. Moreover, we find that high curviness and high numbers of $X^\text{d}$ clusters removed in Step $4$ are more common in CA patterns on tall domains; see Figures~\ref{fig:height}(a)--(b). As we discuss in \S \ref{sec:conclusions}, this has biological implications and may be a natural outgrowth of non-local cell interactions.

\section{Discussion and conclusions}
\label{sec:conclusions}

Different modeling studies offer complementary perspectives and expertise on biological systems; however, as models become more detailed, it is more challenging to understand their behavior---let alone relate their dynamics to the behavior of other models. This is particularly true when stochastic realizations of the same model are variable. To help address this challenge in the case of zebrafish stripe patterns, we developed a method for quantitatively relating two stochastic, microscopic models of cell behavior at large scale. Patterns generated by the models \cite{volkening2018,Owen2020} in our case study are similar by limited observation, but feature important differences in cell density and outlier cells. Our approach uses tools from topological and geometric data analysis (namely persistent homology and $\alpha$-shapes), as well as clustering and data cleaning, to count stripes and interstripes, estimate stripe width, and measure stripe--interstripe boundary curviness. We applied our methodology to the off-lattice model \cite{volkening2018} and the on-lattice model \cite{Owen2020}. These two models \cite{volkening2018,Owen2020}, which are biologically similar but computationally and mathematically different, provide an excellent place to better understand the role of hyper-parameters in our quantification pipeline. We swept across a range of values to choose hyper-parameters that allow for meaningful quantitative descriptions of the models \cite{volkening2018,Owen2020} alongside one another. 

We found that ABM \cite{volkening2018} and CA \cite{Owen2020} patterns are susceptible to different types of noise and ``imperfections". (We use ``imperfections" to broadly refer to patterns with broken stripes or interstripes, spots, or high curviness; because large-scale quantitative data on \emph{in vivo} zebrafish patterns is not available to our knowledge, we do not know if these imperfections are biological errors.) ABM \cite{volkening2018} patterns feature clear separation between stripe and interstripe cells, whereas the CA \cite{Owen2020} often produces stray stripe cells in interstripes, and vice versa. We also observed differences in how ABM and CA patterns break: our methodology suggests that the ABM produces some patterns with breaks in stripes and interstripes. On the other hand, breaks in CA patterns are concentrated in interstripes. Across all of our quantification results, we found that CA \cite{Owen2020} patterns are more variable than ABM \cite{volkening2018} patterns. Interestingly, our methodology uncovers the presence of gold spots in some CA \cite{Owen2020} stripes.

As we discussed in \S \ref{sec:results2}, spots and highly curved interstripes are more common for CA simulations with above-average vertical domain growth. This may be related to the non-local cell behaviors present in the models \cite{volkening2018,volkening2015,Owen2020}: based on empirical findings \cite{Nakamasu}, $M$ cells require signals from interstripe cells at long range for survival. Once stripe width is stretched too far by vertical domain growth in CA patterns, we would expect death of $M$ cells in the middle of the interstripe, followed by appearance of $X^\text{d}$ cells. One possibility is that L-iridophores, an additional cell type that appears later in stripe formation and is not included in the models \cite{volkening2018,Owen2020}, may play a role in pattern maintenance, helping prevent wide stripes from breaking into spots \cite{Frohnhofer,volkening2018}. As empirical data becomes available, better understanding the interplay of domain growth and pattern formation is an exciting direction for future work.

Our study focused on the role of hyper-parameters in interpreting persistent homology and constructing $\alpha$-shapes. We did not sweep across the hyper-parameters involved in Steps $1$ and $2$ in part because computing persistent homology is the main computational bottleneck in our approach. We thus comment on the other hyper-parameters in our quantification pipeline and discuss some alternative choices here. First, cropping ABM and CA domains to the same length and removing the top $10$\% and bottom $10$\% of cells vertically in Step $1$ may have affected our results. It is possible that this could cause some CA patterns with very curvy---but unbroken---(inter)stripes to appear broken on the cropped domains. It is thus worth keeping in mind that our results are descriptions of patterns in a focus domain region.

Second, we did not perform a robustness study of the hyper-parameters related to data cleaning in Step~$2$. For the ABM \cite{volkening2018}, as we noted in \S \ref{sec:step2}, McGuirl \emph{et al.} \cite{McGuirl2020} showed that no cleaning is needed before computing persistent homology, so any reasonable hyper-parameter values (including removing no cells) in Step $2$ works. For CA \cite{Owen2020} patterns, data cleaning is much more important, and selecting the values of hyper-parameters in Step~$2$ is a balance between (1) preserving the structure of CA patterns and (2) removing outlier cells that disrupt our interpretation of persistent homology. The choice of hyper-parameters that we made in \S \ref{sec:step2} allowed us to interpret topological summaries as information about stripes and interstripes in Step $3$. Working with outliers when computing persistent homology is challenging in general, and it would be interesting to apply methods from statistical topological data analysis \cite{Fasy2014,Bendich2011,Chazal2018,Ciocanel2021} to zebrafish patterns in the future.

Third, we made choices about which cell types to base measurements on throughout our quantification process. For example, when characterizing the curviness of stripe--interstripe boundaries, we could have based our measurements on $M$ or $X^\ell$ cells (i.e., curviness of black stripe boundaries), $X^\text{d}$ cells (i.e., curviness of gold interstripe boundaries), or a combination of cells of several types. Like McGuirl \emph{et al.} \cite{McGuirl2020}, we chose to focus on $X^\text{d}$ cells. The overlap of $M$ and $X^\text{d}$ cells at stripe--interstripe boundaries in CA \cite{Owen2020} patterns makes it challenging not only to visually trace out boundaries in CA patterns, but also to define (inter)stripe width. When looking at a pattern like the one in Figure~\ref{fig:sweeping2}(a), where does the top interstripe end and the stripe begin? What cells are part of interstripes, and what cells---if any---are present in spots? Indeed, when observing large sets of simulated patterns from the ABM \cite{volkening2018} and particularly the CA \cite{Owen2020}, we found it very difficult to consistently judge model output visually. This further highlights the value of our automated approach and careful consideration of hyper-parameters when quantifying messy, cell-based patterns at large scale.

Both models \cite{volkening2018,Owen2020} simulate zebrafish mutant patterns, and
McGuirl \emph{et al.} \cite{McGuirl2020} developed methods for quantifying several of these mutants for the ABM \cite{volkening2018}. Future work could build on this approach \cite{McGuirl2020} and our work to quantify CA spot patterns. In terms of (inter)stripe width and curviness, our work with stripe patterns shows that the on-lattice model \cite{Owen2020} produces more variability across simulations than does the off-lattice model \cite{volkening2018}. It would be interesting to investigate whether or not this is also the case for mutant patterns, and to identify what features of the models give rise to this difference in variability. In particular, throughout our work, we did not attempt to determine what causes the differences and similarities that our methods uncover between ABM \cite{volkening2018} and CA \cite{Owen2020} patterns. It may be the structure of stochasticity in cell behaviors, the parameters in model rules, choices of computational implementation, or something else that leads to these features. By developing quantitative methods that can be applied flexibly across models, our work helps opens the door to future studies that address mathematical questions like these and bring the perspectives of different microscopic models together to better understand biological systems.\\ \\

\noindent \textbf{Code availability:}
The MATLAB and Python programs that we developed to quantify on-lattice and off-lattice stripe patterns will be made publicly available on GitLab upon publication.\\

\noindent \textbf{Acknowledgments:} The work of E.C.\ and A.Z.\ was supported in part by the NSF through grant DMS-$1714429$ and by Brown University through Karen T.\ Romer Undergraduate Teaching and Research Awards.
The work of B.S.\ was partially supported by the NSF through grants DMS-$2038039$ and DMS-$2106566$.

\appendix

\section{Additional details on the two models in our case study}\label{sec:app}

To give more intuition into our focal models \cite{volkening2018,Owen2020}, we discuss their rules for two cell dynamics (as examples among many) in more detail here. First, the ABM \cite{volkening2018} and CA \cite{Owen2020} specify that one of the ways that the loose iridophore at position $\textbf{I}^\ell_j$ can swap to dense is when two inequalities are met, namely:
\begin{align}
& \underbrace{\sum_{i=1}^{N_\text{M}} \textbf{1}_{\Omega_\text{loc}^{\textbf{I}^\ell_j}} (\textbf{M}_i)<a \text{~~and~~}
\sum_{i=1}^{N^\text{d}_\text{X}} \textbf{1}_{\Omega_\text{long}^{\textbf{I}^\ell_j}} (\textbf{X}_i^\text{d}) < b}_\text{two conditions must be met} 
\Longrightarrow \text{~~$\textbf{I}^\ell_j$ transforms to dense}, \label{eq:rule2}
\end{align}
where $a=3$ and $b=9$ cells for the ABM \cite{volkening2018}, and $a=1$ and $b=16$ cells for the CA \cite{Owen2020}. Here $N^\text{d}_\text{X}$ and $N_\text{M}$ are the current numbers of $X^\text{d}$ and $M$ cells, respectively. The indicator functions $\textbf{1}_{\Omega_\text{long}^\textbf{z}}(\cdot)$ and $\textbf{1}_{\Omega_\text{loc}^\textbf{z}}(\cdot)$ count the number of cells in a long-range annulus or short-range ball around $\textbf{z}$; see \cite{volkening2018,Owen2020} for length scales. In the ABM, $\textbf{I}^\ell_j$ refers to the coordinates of the $j$th loose iridophore in continuous space, and, in the CA, the coordinates are in discrete space. Similarly, $\textbf{M}_i$ and $\textbf{X}^\text{d}_i$ mark the positions of the $i$th melanophore and dense xanthophore, respectively.

As an example of a dynamic that the models describe in a subtly different way, both models \cite{volkening2018,Owen2020} account for $M$ differentiation from uniformly distributed precursors. Given a randomly selected location $\textbf{z}$ in the domain (either $(x,y)$-coordinates or a grid site), the ABM and CA treat $\textbf{z}$ as the location of a precursor that differentiates into an $M$ cell according to long- and short-range signals. The ABM implements the rule:
\begin{align}
&\underbrace{\sum_{i=1}^{N^\text{d}_\text{X}} \textbf{1}_{\Omega_\text{long}^\textbf{z}} (\textbf{X}_i^\text{d}) + \sum_{i=1}^{N^\text{d}_\text{I}} \textbf{1}_{\Omega_\text{long}^\textbf{z}}(\textbf{I}_i^\text{d}) > \alpha + \beta \sum_{i=1}^{N_\text{M}} \textbf{1}_{\Omega_\text{long}^\textbf{z}} (\textbf{M}_i)}_\text{long-range signals} \text{~~~and} \label{eq:rule1} \\
&\underbrace{\sum_{i=1}^{N^\text{d}_\text{X}} \textbf{1}_{\Omega_\text{loc}^\textbf{z}} (\textbf{X}_i^\text{d}) + \sum_{i=1}^{N_\text{M}} \textbf{1}_{\Omega_\text{loc}^\textbf{z}} (\textbf{M}_i)+  \sum_{i=1}^{N^\text{d}_\text{I}} \textbf{1}_{\Omega_\text{loc}^\textbf{z}} (\textbf{I}_i^\text{d})}_\text{local signals: competition and preventing overcrowding} \le \eta \text{~~} \Longrightarrow \text{~~$M$ differentiates at $\textbf{z}$},\nonumber
\end{align}
where $\alpha = 3$ cells \cite{volkening2015}, $\beta = 3.5$, $\eta = 4$ cells, and $N^\text{d}_\text{I}$ is the number of $I^\text{d}$ cells. Using the same notation, the corresponding CA rule for $M$ differentiation at the randomly selected grid position $\textbf{z}$ is:
\begin{align}
&\underbrace{\sum_{i=1}^{N^\text{d}_\text{X}} \textbf{1}_{\Omega_\text{long}^\textbf{z}} (\textbf{X}_i^\text{d}) + \sum_{i=1}^{N^\text{d}_\text{I}} \textbf{1}_{\Omega_\text{long}^\textbf{z}} (\textbf{I}_i^\text{d}) > \alpha + \beta \sum_{i=1}^{N_\text{M}} \textbf{1}_{\Omega_\text{long}^\textbf{z}} (\textbf{M}_i)}_\text{long-range signals (same as Eqn. \eqref{eq:rule1})} \text{~~~and} \label{eq:rule1c}\\
&\underbrace{\sum_{i=1}^{N^\text{d}_\text{X}} \textbf{1}_{\Omega_\text{loc}^\textbf{z}} (\textbf{X}_i^\text{d}) \le \gamma \sum_{i=1}^{N_\text{M}} \textbf{1}_{\Omega_\text{loc}^\textbf{z}} (\textbf{M}_i) \text{~and~}  \sum_{i=1}^{N^\text{d}_\text{I}} \textbf{1}_{\Omega_\text{loc}^\textbf{z}} (\textbf{I}_i^\text{d})\le \kappa}_\text{local signals: competition}  \text{~~} \Longrightarrow \text{~~$M$ differentiates at $\textbf{z}$}, \nonumber
\end{align}
where $\alpha = 3$ cells, $\beta = 10$, $\gamma = 12$, and $\kappa = 3$ cells. The difference in $\beta$ for the ABM and CA long-range terms in Eqn.~\eqref{eq:rule1} and Eqn.~\eqref{eq:rule1c} may be related to the difference in $M$ density in these models; $\alpha$ is as delay term, slowing how quickly $M$ cells differentiate after the appearance of interstripe cells.

\section{Supplementary Material} \label{sec:supp}

\begin{figure}[h]
\centering
\includegraphics[width=0.9\textwidth]{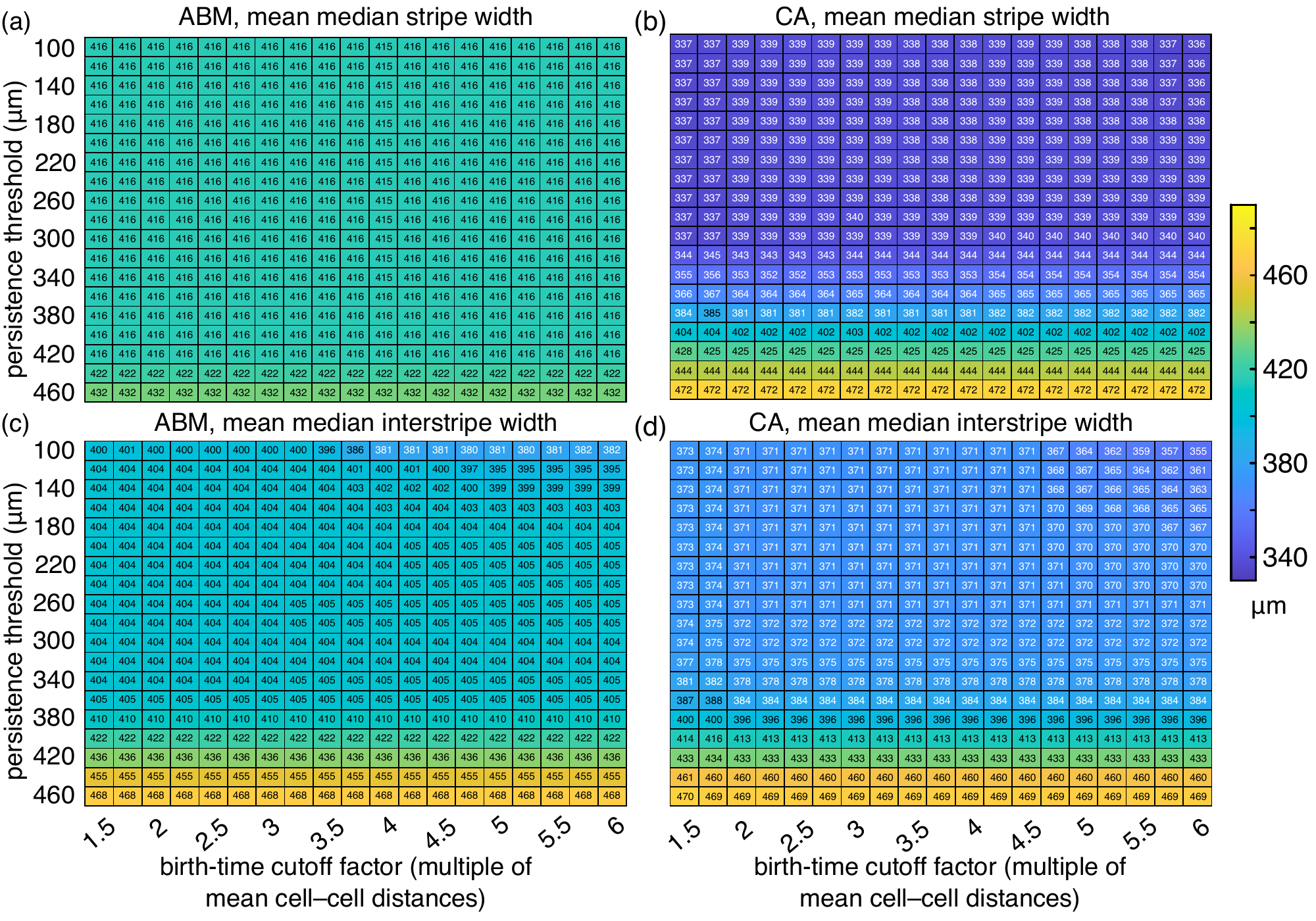}
\caption{Effects of birth-time cutoff $B_m$ and persistence threshold $P_m$ on quantifying stripe and interstripe width in Step $3$. Results are based on $100$ ABM \cite{volkening2018} and $100$ CA \cite{Owen2020} patterns, cropped and cleaned as described in Steps $1$ and $2$, that we quantify under each choice of $(B_m,P_m)$ given. Here $B_m$ is a factor multiplied by the appropriate mean cell--cell distance. We compute the median (inter)stripe width per simulation and then take the mean across $100$ simulations. For (a) the ABM \cite{volkening2018} and (b) the CA \cite{Owen2020}, stripe width is insensitive across many values of our hyper-parameters. For (c) ABM and (d) CA interstripe width, we see similar insensitivity to hyper-parameters.  }
\end{figure}

\bibliographystyle{plain}
\bibliography{references}
\end{document}